\documentclass[%
 notitlepage,
 twocolumn,
superscriptaddress,
nofootinbib,
amsmath,amssymb,
aps,
pra,
longbibliography,
]{revtex4-2}
\usepackage[T1]{fontenc}
\usepackage[latin9]{inputenc}
\usepackage{float}
\usepackage{textcomp}
\usepackage{mathtools}
\usepackage{bbold}
\usepackage{cancel}
\usepackage{soul}
\usepackage{amsfonts}
\usepackage{graphicx}
\usepackage{physics}
\usepackage{hyperref}
\hypersetup{
citecolor=blue,
colorlinks=true,
urlcolor=magenta,
}
\usepackage{ulem}

\usepackage{xcolor}

\usepackage{comment}

\newcommand{\VILNIUS}{Institute of Theoretical Physics and Astronomy,
	Vilnius University, Saul\.etekio 3, LT-10257, Vilnius, Lithuania}

\newcommand{\IOPPAS}{Institute of Physics PAS, Aleja Lotnik\'ow 32/46, 02-668 Warszawa, Poland}

\newcommand{\LPL}{Laboratoire de Physique des Lasers, CNRS UMR 7538, Universit\'e Sorbonne Paris Nord, F-93430, Villetaneuse, France}

\def\Isq{\bar{i}_I^2}
\def\Iqu{\bar{i}_I^4}

\begin{document}

\title{Effective light-induced Hamiltonian for atoms with large nuclear spin
}

\author{D. Burba}
\affiliation{\VILNIUS}
\author{H. Dunikowski}
\affiliation{\IOPPAS}
\author{M. Robert-de-Saint-Vincent}
\affiliation{\LPL}
\author{E. Witkowska}
\affiliation{\IOPPAS}
\author{G. Juzeli\=unas}
\affiliation{\VILNIUS}

\date{\today}
	
\begin{abstract}
Ultra-cold fermionic atoms, having two valence electrons, exhibit a distinctive internal state structure, wherein the nuclear spin becomes decoupled from the electronic degrees of freedom in the ground electronic state. Consequently, the nuclear spin states are well isolated from the environment, rendering these atomic systems an opportune platform for quantum computation and quantum simulations.
Coupling with off-resonance light is an essential tool to selectively and coherently manipulate the nuclear spin states. 
In this paper, we present a systematic derivation of the effective Hamiltonian for the nuclear spin states of ultra-cold fermionic atoms due to such an off-resonance light. We obtain compact expressions for the scalar, vector and tensor light shifts taking into account both linear and quadratic contributions to the hyperfine splitting. The analysis has been carried out using the Green operator approach and solving the corresponding Dyson equation. 
Finally, we analyze different scenarios of light configurations which lead to the vector- and tensor-light shifts, as well as the pure spin-orbit coupling for the nuclear spin.
\end{abstract}

\maketitle

\section{Introduction}
Alkaline earth atoms stand as a prime platform for the realms of quantum simulation and computation, as well as for high-performance optical clocks~\cite{PhysRevLett.101.170504, Gonz_lez_Cuadra_2023, PhysRevResearch.5.043175,PhysRevLett.102.110503,PhysRevX.13.041035,Deutsch_2021,Mamaev2022}.
These atoms have two valence electrons forming a closed shell $ns^2$ with zero spin and angular momentum quantum numbers ($S= L=J=0$) in the electronic ground state $^1S_0$. 
They are also characterized by forbidden optical transitions to electronic triplet states $^3P_{j}$, with $j=0,1,2$. Fermionic isotopes possess a nuclear spin that is highly isolated from the environment rendering them a convenient platform for the applications mentioned above~\cite{PhysRevLett.101.170504, Gonz_lez_Cuadra_2023, PhysRevResearch.5.043175,PhysRevLett.102.110503,PhysRevX.13.041035,Deutsch_2021,Mamaev2022}.
The most notable alkaline-earth species are Mg, Ca, Sr, but other species - in particular Yb, and the group IIB transition metals (Hg \cite{McFerran2012},  Cd \cite{Yamaguchi2019}, Zn\,\cite{Wang2007Zn}) share these properties\,\cite{Derevianko2011, Dzuba2019,Ludlow2015RMP}. All of them but Zn have been laser-cooled and trapped in the ground state. It is noteworthy that the number of nuclear spin states of their fermionic isotopes can be quite large, such as $N=10$ for $^{87}$Sr, providing beneficial features useful for quantum simulation and computation~\cite{Deutsch_2021}. The nuclear spin in these atoms thus represents a convenient carrier of quantum information taking into account the long coherence times and the ability of coherent spin control with magnetic and optical fields. 

Manipulation of atomic spin states by off-resonant laser fields was studied extensively in the context of both the alkali \cite{Deutsch_2007, PhysRevA.79.013404, Deutsch_2010, PhysRevA.79.013404,LeKien2013,Goldman:2014RPP} and alkaline earth atoms \cite{Deutsch_2021,Mamaev2022}. The light-induced effective atomic spin Hamiltonian can be generally cast in terms of components known as the scalar, vector and tensor light shifts \cite{Deutsch_2010, LeKien2013}. 
%
%

Typically, the evaluation of these light shifts involves numerical summation over transitions to various excited states, with differing strengths expressed using e.g. Clebsch-Gordan coefficients~\cite{Deutsch_2010, LeKien2013}.
Simple analytical forms of the effective Hamiltonian and the scalar, vector and tensor components of the light shifts are desired to ease practical experimental implementations and theoretical understanding.

Here we develop a formalism allowing us to obtain compact analytical expressions for the scalar, vector and tensor light shifts of the nuclear spin states for the alkaline-earth-like atoms. We find that the three light shifts are simple rational functions expressed solely in terms of the nuclear spin quantum number, hyperfine splitting energies, and detuning. The method involves the Green operator (the resolvent) approach and the solution of the corresponding Dyson equation, allowing to by-pass explicit analysis of transitions to the specific hyperfine states of the excited state manifold. The technique is related to the one used to analyze the vector light shifts of the alkali atoms presented in Sec.~4 of Ref.~\cite{Goldman:2014RPP}. Yet, there are significant differences between the two treatments. In the present analysis of the alkaline-earth-like atoms the electron spin is zero in the ground state, and it is the hyperfine splitting of the excited electronic states which provides the vector and tensor light shifts for the (nuclear) spin states of the ground state atoms. On the other hand, for alkali atoms with a single unpaired outer electron, it is the excited state fine structure splitting which is responsible for the vector light shifts for the spin states in the ground electronic manifold \cite{Goldman:2014RPP}.
In our analytical calculations, we have managed to include not only linear but also quadratic terms of the hyperfine coupling operator. Therefore the theory provides a very accurate description of the light shifts for the alkaline-earth-like atoms. To resolve the hyperfine splitting without significant perturbation by spontaneous emission we consider the light shifts due to the $^3P_1$ intercombination line characterized by a relatively large frequency of the hyperfine splitting $A_{\rm HF}$ and a comparatively small spontaneous emission rate $\Gamma$. As an example, for $^{87}$Sr atoms $A_{\rm HF}/2\pi$ is  around $2.6 \times 10^{5}$ kHz, whereas $\Gamma/2\pi$ is of the order of 10 kHz~\cite{Miyake2019} for the $^3P_1$ line. This allows an efficient light-induced manipulation of the nuclear spin states with minimum decoherence. In the subsequent text, we will consider $^{87}$Sr as a typical example. Nevertheless, our analytical findings are broad-reaching and can be applied to other alkaline earth (or alkaline earth-like) atoms with the same structure.

The paper is organised as follows. In Section \ref{sec:one} we introduce the fine and hyperfine structures of alkali earth or alkali earth-like atoms. Next, in Section \ref{sec:two}, we present our theory for the derivation of effective Hamiltonian describing the spin ground-state manifold when an atom is coupled with laser radiation. In Section \ref{sec:three}, we discuss typical examples of light configuration and the resulting atom-light coupling Hamiltonian, as well as prospects for the realization of spin-orbit coupling for fermionic atoms with nuclear spin. Finally, the concluding Section~\ref{sec:four} summarizes our findings.

\section{Manifolds of ground and excited states for alkaline-earth-like atoms}
\label{sec:one}

We will consider the optical control of nuclear spin for alkali earth or alkali earth-like atoms with two $s$-shell valence electrons characterised by a nuclear spin $i_I$. We will study the fermionic isotopes which have non-zero nuclear spin. Bosonic alkaline-earth-like atoms possess zero nuclear spin, as they contain an even number of protons and even number of neutrons. Thus the bosonic species are not of interest for the present analysis of the manipulation of atomic nuclear spin. 

In the ground electronic manifold denoted by $^{1}S_{0}$ the spins of outer electrons of alkali earth-like atoms form a singlet, so the quantum numbers of the electronic spin ${\bf S}$ and the orbital angular momenta ${\bf L}$ are zero, as well as the total electronic momentum ${\bf J} = {\bf L} + {\bf S}$, namely $s=l=j=0$. 
It is a reason why the fine and hyperfine couplings are not affecting the ground
electronic state. The energy of this state is taken to be zero $E_{g}=0$. 

The excited electronic state of interest $|j,m_{j}\rangle $ with $j=0,1,2$ belongs to the manifold $^{3}P_{j}$ where spins of the outer electrons form a triplet, and the state of the atom is characterized by the spin and orbital angular momentum quantum numbers, $s=1$  and $l=1$, respectively~\footnote{Alternatively, the theory presented here is also applicable for virtual transitions via the $^{1}P_1$ manifold. However, we will concentrate on the $^{3}P_1$ manifold, as its higher ratio between hyperfine structure and linewidth makes it more relevant experimentally due to much smaller rate of spontaneous emission.}.
The state $|j,m_{j}\rangle $ is an eigenstate of $\mathbf{S}^{2}$,
$\mathbf{L}^{2}$, $\mathbf{J}^{2}$ and $\mathbf{J}_{z}$ with the corresponding eigenvalues
$\hbar^{2}s\left(s+1\right)=2\hbar^{2}$, $\hbar^{2}l\left(l+1\right)=2\hbar^{2}$,
$\hbar^{2}j\left(j+1\right)$ and $\hbar m_{j}$. Without including the fine and hyperfine couplings the excited state is degenerate concerning $j$. The fine structure coupling removes this degeneracy.

The fine structure (FS) coupling between the total electron spin and orbital angular momentum is described by the Hamiltonian
$H_{\rm FS}=\frac{A_{\rm FS}}{\hbar}\mathbf{L}\cdot\mathbf{S}=\frac{A_{\rm FS}}{\hbar}\left(\mathbf{J}^{2}-\mathbf{L}^{2}-\mathbf{S}^{2}\right)/2$ providing $j$-dependent energies 
$E_{\rm FS}^{\left(j\right)}=\hbar A_{\rm FS}\left[j\left(j+1\right)-4\right]/2$
for $l=s=1$. Therefore, one has
\begin{equation}
E_{\rm FS}^{\left(j=0\right)}=-2\hbar A_{\rm FS}\,,\,\, 
E_{\rm FS}^{\left(j=1\right)}=-\hbar A_{\rm FS}\,,\,\, 
E_{\rm FS}^{\left(j=2\right)}=\hbar A_{\rm FS}\,.
\label{eq:E_fs^j-1}
\end{equation}
The state with $j=2$ is not accessible via the dipole transitions
from the ground state manifold $^{1}S_{0}$ with $j=0$, and thus will not be considered by us in the further part of the text. 
The transition to the state
with $j=0$, also known as "clock transition", is double forbidden, both by spin and orbital degrees of freedom. It is very weak, and only relevant if the radiation frequency is very close to this state's resonance.
Therefore, in what follows, we will consider only the transition to the excited state manifold $^{3}P_{1}$ corresponding to $j=1$. This transition called the intercombination line, is forbidden by spin flip, but opens due to $j-j$ coupling admixing with the $^1P_1$ state~\cite{Foot2004}.

The hyperfine (HF) coupling between the
total electronic angular momentum $\bf{J}$ and the nuclear spin $\bf{I}$ is generally given by the Hamiltonian (see e.g. ref.~\cite{LeKien2013}):
\begin{equation}
H_{\mathrm{HF}}=\frac{A_{{\rm HF}}^{\prime}}{\hbar}\mathbf{I}\cdot\mathbf{J}+B_{{\rm HF}}\frac{6\left(\mathbf{I}\cdot\mathbf{J}\right)^{2}+3\hbar^{2}\mathbf{I}\cdot\mathbf{J}-2\mathbf{I}^{2}\mathbf{J}^{2}}{4\hbar^{3}i_{I}(2i_{I}\text{\textminus}1)j(2j\text{\textminus}1)}\,,\label{eq:H_hf-0}
\end{equation}
where in the present situation $j=1$. Here $A_{{\rm HF}}^{\prime}$ and $B_{{\rm HF}}$ are HF constants. In the case of strontium atoms, their values are $A_{{\rm HF}}^{\prime}/ 2 \pi=- 260085 \pm 2\,$kHz and $B_{{\rm HF}}/ 2 \pi=-  35667 \pm 21\,$kHz~\cite{Miyake2019,Putlitz63}. It is convenient to represent $H_{\mathrm{HF}}$ as
\begin{equation}
H_{\mathrm{HF}}=\frac{A_{{\rm HF}}}{\hbar}\left[\mathbf{I}\cdot\mathbf{J}+\frac{\gamma}{\hbar^{2}}\left(\mathbf{I}\cdot\mathbf{J}\right)^{2}\right]\,,\label{eq:H_hf}
\end{equation}
where $A_{\rm HF}=A_{{\rm HF}}^{\prime} + 3 B_{{\rm HF}}/[4 i_I (2 i_I -1)j (2j-1)]$, and the dimensionless parameter $\gamma=6 B_{{\rm HF}}/[A_{{\rm HF}}4 i_I (2 i_I -1)j (2j-1)]$ describing a relative strength of the quadratic term is generally much less than the unity. For $^{87}$Sr atoms one has $\gamma \approx 0.006$, and we will use this value in specific examples. In Eq.~(\ref{eq:H_hf}) a uniform shift of energy proportional to $\mathbf{I}^{2}\mathbf{J}^{2}$
 has been incorporated to the excitation energy $E_{\rm FS}^{\left(j=1\right)}$, whereas the
two terms of Eq.~(\ref{eq:H_hf-0}) proportional to $\mathbf{I}\cdot\mathbf{J}$
have been merged to a single term characterised by the modified frequency $A_{{\rm HF}}$.

\begin{figure}
\centering
\includegraphics[width=0.8\linewidth]{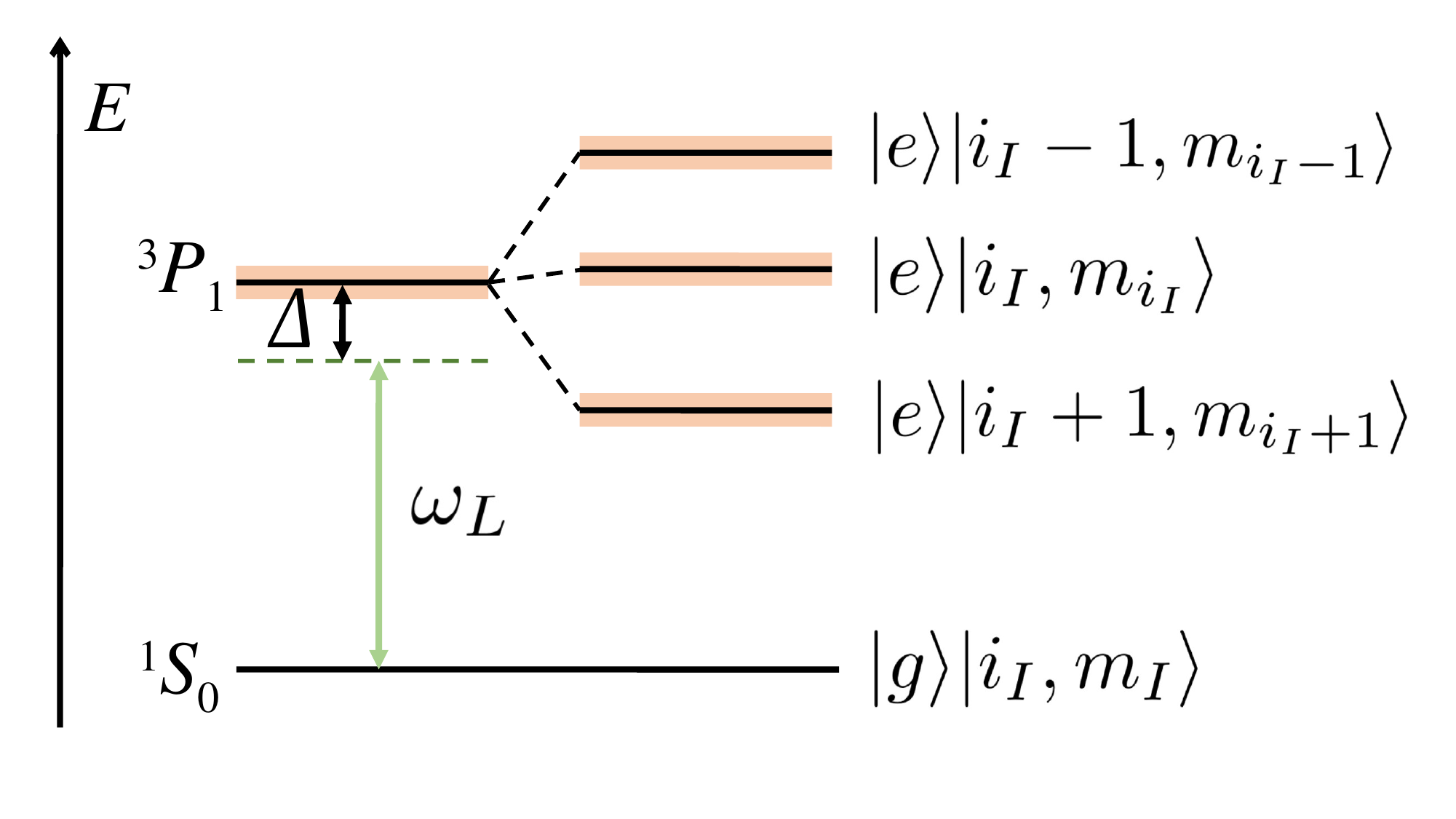}
\caption{
A sketch of the atomic levels for alkaline-earth-like atoms illustrating the hyperfine structure of the excited electronic line $^3P_1$. The order of the excited states depends on the sign of $A_{\rm HF}$; here we take $A_{\rm HF} <0$ as in $^{87}$Sr. The probe frequency $\omega_L$ is tuned in a vicinity of the transition $^{1}S_0 \to {}^3P_1$ manifold with the detuning $\Delta$.
}
\label{fig:fig1}
\end{figure}

The eigenstates of hyperfine Hamiltonian are also eigenstates of $\mathbf{F}^{2}$ ($\mathbf{F} = \mathbf{I} + \mathbf{J}$),
$\mathbf{I}^{2}$, $\mathbf{J}^{2}$ and $\mathbf{F}_z$, and the corresponding eigenvalues are
$\hbar^{2}f\left(f+1\right)$, $\hbar^{2}i_I\left(i_I+1\right)$,
$\hbar^{2}j\left(j+1\right)$, 
and $\hbar m_f$. For the excited state $^3P_1$ manifold of interest one has $j=1$, 
so $f$ can take the values $f=i_{I}$ and $f=i_{I}\pm1$.
The energies of these excited states are: 
\begin{align}
E_{i_{I}+1}&=\hbar A_{{\rm HF}}\, i_{I}\left(1+\gamma i_{I}\right)\,,\label{eq:E_i_plus_with-quadr-term}\\ 
E_{i_{I}}&=-\hbar A_{{\rm HF}}\left(1-\gamma\right)\,,\label{eq:E_i_with-quadr-term}\\
E_{i_{I}-1}&=-\hbar A_{{\rm HF}}\left(i_{I}+1\right)\left[1-\gamma\left(i_{I}+1\right)\right]\,.\label{eq:E_i_minus_with-quadr-term}
\end{align}
The hyperfine structure for the electronic line $^3P_1$ is schematically represented in Fig.~\ref{fig:fig1}. 

\section{Effective Hamiltonian}
\label{sec:two}

We consider the effects of light field $\mathbf{E}$ on atomic nuclear spin states in the ground electronic manifold $^{1}S_{0}$ due to the virtual transition to the excited state manifold $^{3}P_{1}$.
In the dipole approximation, the interaction between a monochromatic light field characterized by the electric field strength $\mathbf{E}$ and an atom can be described by the operator
\begin{equation}
    V= - \mathbf{d} \cdot\mathbf{E}= -\sum_q {d}_q \tilde{\mathcal{E}}_q \cos(\phi_q -\omega_L t). \label{eq:V}
\end{equation}
where the components of the electric field amplitude $\tilde{\mathcal{E}}_q$ and phase $\phi_q$ of the radiation are generally position dependent. Here ${\bf d}=-e \sum_\alpha {\bf r}_\alpha$ is the electric dipole operator, $e$ is an electron charge and ${\bf r}_\alpha$ is the position of the $\alpha$th electron within the atom. 

We will go to the frame rotated at the driving frequency $\omega_L$ and subsequently apply the rotating wave approximation (RWA) by removing terms oscillating at frequencies $\omega_L$, $2 \omega_L$ in the transformed interaction Hamiltonian which then takes the form, see e.g. Chapter 4 of Ref.~\cite{Goldman:2014RPP}:
\begin{equation}
    V = - \frac{1}{2} \sum_q \left[ \mathcal{E}^*_q P_g d_q P_e + \mathcal{E}_q P_e d_q P_g  \right],
\end{equation}
where we introduced the complex electric field
$\mathcal{E}_q = \tilde{\mathcal{E}}_q e^{i \phi_q}$.

In the above formula,
$P_{g}$ and $P_{e}$ are projector operators onto the ground and excited
state manifolds $^{1}S_{0}$ and $^{3}P_{1}$, respectively.

\subsection{Analysis of effective Hamiltonian}

If the incident light $\boldsymbol{\mathcal{E}}$ with components $\mathcal{E}_q$ is sufficiently weak and/or well detuned from atomic resonances, the saturation parameter is small, and the atom-light interaction effectively provides coupling between the nuclear spin states $ |i_I, m_{i_I}\rangle$ of the electronic ground state $|g\rangle $ via virtual transitions to the excited states without populating them\footnote{This means we are not considering the resonant regime ($|\Delta - E_f| \lessapprox |d_{ge}\mathbf{E}|$), where $E_f$ is energy of hyperfine state, given by Eqs.~(\ref{eq:E_i_plus_with-quadr-term})-(\ref{eq:E_i_minus_with-quadr-term}) and $|d_{ge}|=\sqrt{|d_{ge}^2|}$ is defined in Eq.~(\ref{eq:d^2}).
In the resonant situation one would have a significant excited state population and observe Rabi oscillations.}. Specifically, the adiabatic elimination of the excited states can be described in the second order (with respect to the interaction $V$) via Schrieffer-Wolff transformation, see e.g. Supplementary Materials of~\cite{Hern_ndez_Yanes_2022}. This provides the corresponding second order effective atomic Hamiltonian for the ground state manifold:
\begin{equation}
    \hat{H}_{\rm eff} = P_g V   G   V P_g 
\label{H_eff-definition}
\end{equation}
where
\begin{equation}
    G=(\Delta - H_{\rm HF})^{-1}
\label{eq:G-definition}
\end{equation}
is the Green operator
and $\Delta=E_{\rm FS}^{(j=1)}-E_g-\hbar \omega_L$ is the detuning. 
Although our analysis is perturbative to second order in electron-light interaction $V$, the hyperfine interaction $H_{\rm HF}$ is treated exactly. Note that by adding an imaginary part to the detuning $\Delta$, one can include effects of losses due to spontaneous emission. This is discussed in Sec.~\ref{losses}.

The effective atomic Hamiltonian can be expressed as 
\begin{equation}
    \hat{H}_{\rm eff}  =\frac{1}{4}\mathcal{E}_{s}^{*}D_{s,q}\mathcal{E}_{q}.
    \label{eq:H^eff'}
\end{equation}
Here and below the summation over repeated indices of vectors and tensors is assumed. In Eq.\eqref{eq:H^eff'} we introduced the tensor operator acting on the ground state manifold:
\begin{align}
D_{s,q} & =P_{g}d_{s}P_e G P_e d_{q}P_{g}\,,\label{eq:D_ij}
\end{align}
where the Green operator $G$ obeys a Dyson-type equation (identity) 
\begin{align}
G & =\frac{1}{\Delta}+\frac{1}{\Delta}H_{\mathrm{HF}}G,
\label{eq:Dyson-identity-for-G}
\end{align}
see Appendix~\ref{app:Dyson}.
Combining Eqs.~ \eqref{eq:D_ij} and \eqref{eq:Dyson-identity-for-G}, one has
\begin{align}
D_{s,q} & =\frac{\left|d_{ge}^{2}\right|}{\Delta}
P_{g}
\delta_{sq}
+ \frac{1}{\Delta}
P_{g}d_{s} P_e H_{\rm HF} G P_{e}d_{q}P_{g}\,,
\label{eq:D_ij-equation-0}
\end{align}
where
\begin{equation}
\left|d_{ge}^{2}\right|=\frac{1}{3}\left\langle g\right|d_{q}P_{e}d_{q}\left|g\right\rangle \label{eq:d^2}
\end{equation}
characterizes the strength of the dipole transition and equals to the third of the reduced matrix element
$\langle g|d_{q}P_{e}d_{q}|g\rangle = \sum_e  |\langle e |d_{q}|g\rangle|^2$. The dipole transition is weak, but non-zero. 
This is because the excited state manifold $^{3}P_{1}$
contains a small admixture of the manifold $^{1}P_{1}$ with
the same $j=1$, but zero spin $s=0$, and the transitions to the latter manifold $^{1}S_{0}\rightarrow{}^{1}P_{1}$ are dipole allowed.

According to the Land\'e projection theorem~\cite{Deutsch_2010}, the effective Hamiltonian (\ref{eq:H^eff'}) can be represented in the form
\begin{align}
    \hat{H}_{\rm eff} & = 
    \frac{b_0}{4}\left(\boldsymbol{\mathcal{E}}^{*}\cdot\boldsymbol{\mathcal{E}}\right)
    +
    i\frac{b_1}{4\hbar}
\textbf{\ensuremath{\mathbf{I}}}\cdot\left(\boldsymbol{\mathcal{E}}^{*}\times\boldsymbol{\mathcal{E}}\right)\nonumber \\
&+
\frac{b_{2}}{4\hbar^2}
\left[ 
\left(\boldsymbol{\mathcal{E}}^{*}\cdot\textbf{\ensuremath{\mathbf{I}}}\right)\left(\boldsymbol{\mathcal{E}}\cdot\textbf{\ensuremath{\mathbf{I}}}\right)
-\frac{1}{3}\left|\boldsymbol{\mathcal{E}}\right|^{2}\mathbf{I}^2 + {\rm h.c.}
\right].
\label{eq:generalDij}
\end{align}
where the coefficients $b_0,b_1$ and $b_2$ characterize the strengths of the scalar, vector and tensor terms. Conventionally the effective Hamiltonian is calculated using the explicit expressions for the transition matrix elements involving the Clebsch-Gordon coefficients. 
Here, we perform analytical calculations of the coefficients $b_0,b_1$ and $b_2$ defining the effective Hamiltonian to derive their simple forms in a self-consistent way 
based on solution of Eq.~(\ref{eq:D_ij-equation-0}) for $D_{s,q}$.
To this end, it is convenient to introduce a vector operator $\mathbf{K}$ with Cartesian components derived from $D_{s,q}$: 
\begin{equation}
K_{s}=D_{s,q}\mathcal{E}_{q}\,.\label{eq:F_i_main}
\end{equation}
In terms of $\mathbf{K}$, the effective Hamiltonian of Eq~\eqref{eq:H^eff'} reads:
\begin{equation}
\hat{H}_{{\rm eff}} =
\frac{1}{4}\boldsymbol{\mathcal{E}}^{*}\cdot\mathbf{K}\,.\label{eq:H_eff-F}
\end{equation}
Multiplying Eq.~\eqref{eq:D_ij-equation-0} by $\mathcal{E}_{q}$ from the right-hand side, we write down the relation for the vector-operator $\mathbf{K}$
\begin{align}
\mathbf{K} & =\frac{1}{\Delta}P_{g}\left|d_{ge}^{2}\right|\mathcal{\boldsymbol{\mathcal{E}}}\nonumber \\
&+\frac{A_{{\rm HF}}}{\hbar\Delta}P_{g}\mathbf{d}\left(\mathbf{I}\cdot\mathbf{J}\right)\left(1+\frac{\gamma}{\hbar^{2}}\mathbf{I}\cdot\mathbf{J}\right)GP_{e}\left(\mathbf{d}\cdot\boldsymbol{\mathcal{E}}\right)P_{g}\,.\label{eq:K-equation}
\end{align}
As demonstrated in Appendix~\ref{subsec:Append:Equation for K}, this provides the following equation for $\mathbf{K}$:
\begin{align}
\eta\mathbf{K} & =\frac{1}{\Delta}P_{g}\left|d_{ge}^{2}\right|\boldsymbol{\mathcal{E}}+i\frac{A_{{\rm HF}}}{\Delta}\left(1-\gamma\right)\left(\textbf{\ensuremath{\mathbf{I}}}\times\mathbf{K}\right)-
\nonumber \\
&-\gamma\frac{A_{{\rm HF}}}{\hbar\Delta}\mathbf{I}\left(\textbf{\ensuremath{\mathbf{I}}}\cdot\mathbf{K}\right)\,,
\label{eq:K-vector-equation}
\end{align}
where $\eta=1-\gamma A_{{\rm HF}}\hbar i_{I}\left(i_{I}+1\right)/\Delta$.

Anticipating the effective Hamiltonian of the form~(\ref{eq:generalDij}), we will look for the following ${\bf K}$ ansatz
\begin{align}
\mathbf{K} & =\left[a_{0}\boldsymbol{\mathcal{E}}
-i\frac{a_{1}}{\hbar}\textbf{\ensuremath{\mathbf{I}}}\times\boldsymbol{\mathcal{E}}
+\frac{a_{2}}{\hbar^{2}}\textbf{\ensuremath{\mathbf{I}}}\left(\textbf{\ensuremath{\mathbf{I}}}\cdot\boldsymbol{\mathcal{E}}\right)\right]P_{g},
\label{eq:F-vector-solution}
\end{align}
where $a_{0}$, $a_{1}$ and $a_{2}$ are real coefficients. Inserting the ansatz~(\ref{eq:F-vector-solution}) into Eq.~(\ref{eq:K-vector-equation}), one gets a system of equations for these coefficients, giving  
\begin{align}
a_{0}&=
-\frac{|d_{ge}^{2}|}{\hbar A_{\rm HF}}
\frac{
\bar{E}^{(+)}
}
{\left(\bar{\Delta}-\bar{E}_{i_{I}+1}\right)\left(\bar{\Delta}-\bar{E}_{i_{I}-1}\right)}\,,
\label{eq:c_0-result-with-quadr-term}
\\
a_{1}&=
\frac{|d_{ge}^{2}|}{\hbar A_{\rm HF}}
\frac{\bar{E}_{i_I}}
{\left(\bar{\Delta}-\bar{E}_{i_{I}+1}\right)\left(\bar{\Delta}-\bar{E}_{i_{I}-1}\right)}\,,
\label{eq:c_1-result-with-quadr-term}
\\
a_{2}&=
\frac{|d_{ge}^{2}|}{\hbar A_{\rm HF}}
\frac{
\gamma \bar{E}^{(-)} - \bar{E}_{i_{I}}^2 
}
{\left(\bar{\Delta}-\bar{E}_{i_{I}+1}\right)\left(\bar{\Delta}-\bar{E}_{i_{I}}\right)\left(\bar{\Delta}-\bar{E}_{i_{I}-1}\right)}\,,
\label{eq:c_2-result-with-quadr-term}
\end{align}
with
\begin{equation}
\bar{E}^{(\pm)}=(\bar{E}_{i_{I}+1}\pm\bar{E}_{i_{I}}+\bar{E}_{i_{I}-1})/2 - \bar{\Delta}.
\label{eq:E^pm}
\end{equation}
where $\bar{\Delta}=\Delta/(\hbar A_{\rm HF})$ and $\bar{E}_f = E_f/(\hbar A_{\rm HF})$ (with $f=i_{I} - 1, i_{I}, i_{I}+1 $) are, respectively, the dimensionless detuning and energies of hyperfine states, and $E_{f}$ is defined by Eqs.~\eqref{eq:E_i_plus_with-quadr-term}-\eqref{eq:E_i_minus_with-quadr-term}. 
Details of derivation are in Appendix~\ref{Subsec:Solution of equation for K}.

Substituting the solution~\eqref{eq:F-vector-solution} for $\mathbf{K}$ into Eq.~\eqref{eq:H_eff-F}, one arrives at the explicit result for the ground state effective Hamiltonian expressed in terms of the coefficients $a_{0,1,2}$
\begin{align}
    \hat{H}_{\rm eff} & = 
    \frac{a_0}{4}\left(\boldsymbol{\mathcal{E}}^{*}\cdot\boldsymbol{\mathcal{E}}\right)
    +
    i\frac{a_1}{4\hbar}
\textbf{\ensuremath{\mathbf{I}}}\cdot\left(\boldsymbol{\mathcal{E}}^{*}\times\boldsymbol{\mathcal{E}}\right) + \nonumber \\
&+
\frac{a_{2}}{4\hbar^2}
\left(\boldsymbol{\mathcal{E}}^{*}\cdot\textbf{\ensuremath{\mathbf{I}}}\right)
\left(\boldsymbol{\mathcal{E}}\cdot\textbf{\ensuremath{\mathbf{I}}}\right)\,,
\label{eq:general-H_eff-c-coeff}
\end{align}
where for brevity we have omitted the ground state projector operator $P_{g}$. 

Note that the self-consistent method we used to find the vector ${\bf K}$ of Eq.~(\ref{eq:F-vector-solution}) can be applied on the level of the tensor $D_{s,q}$ featured in Eqs.~(\ref{eq:D_ij}) and (\ref{eq:D_ij-equation-0}). In that case one finds
\begin{equation}
D_{s,q}=a_{0}\delta_{sq}+i\frac{a_{1}}{\hbar}I_{k}\epsilon_{ksq}+\frac{a_{2}}{\hbar^{2}}I_{s}I_{q}\,.
\end{equation}
This leads to the same form of the effective Hamiltonian~(\ref{eq:general-H_eff-c-coeff}) with the same coefficients $a_{0,1,2}$ defined by Eqs.~(\ref{eq:c_0-result-with-quadr-term})-(\ref{eq:c_2-result-with-quadr-term}).

Using the fact that 
$\boldsymbol{\mathcal{E}}^{*}
\cdot
\left(\boldsymbol{I}
\times
\boldsymbol{\mathcal{E}}\right)
=
-\boldsymbol{\mathcal{E}}^{*}\cdot\left(\boldsymbol{\mathcal{E}}
\times
\boldsymbol{I}\right)
=-
\left(\boldsymbol{\mathcal{E}}^{*}
\times
\boldsymbol{\mathcal{E}}\right)\cdot\boldsymbol{I}$,
we express the effective Hamiltonian (\ref{eq:general-H_eff-c-coeff}) in the symmetric and traceless form as in Eq.~(\ref{eq:generalDij}). Consequently, we obtain the following relations for the coefficients $b_{0}$, $b_1$ and $b_2$ defining the scalar, vector and tensor light shifts
\begin{equation}
b_{0}=a_{0}+ \Isq a_2/3,\,\,
b_{1}=a_{1}+a_{2}/2\,\, \mathrm{and}\,\,
b_{2}=a_{2}/2.
\label{b-vs-a}
\end{equation}
with
\begin{equation}
\Isq \equiv \mathbf{I}^2/\hbar^2= i_I(i_I+1)\,,
\label{eq:i(i+1)-def}
\end{equation}
It is convenient to define the dimensionless $b$ coefficients 
\begin{equation}   
\bar{b}_{u} = b_{u} \hbar A_{\rm HF} /|d_{ge}^2|\,\,,\quad \mathrm{with} \quad u=0,1,2\,.
\end{equation}

Calling on Eqs.~\eqref{eq:c_0-result-with-quadr-term}-\eqref{eq:c_2-result-with-quadr-term} and \eqref{b-vs-a}, one arrives at the explicit expressions for  the latter coefficients:
\begin{align}
\bar{b}_{0}&=
\frac{
-3\bar{E}^{(+)}
\left(\bar{\Delta}-\bar{E}_{i_{I}}\right)
+
\Isq
\left(\gamma \bar{E}^{(-)}
 - \bar{E}_{i_{I}}^2
\right)
}
{3\left(\bar{\Delta}-\bar{E}_{i_{I}+1}\right)
\left(\bar{\Delta}-\bar{E}_{i_{I}}\right)
\left(\bar{\Delta}-\bar{E}_{i_{I}-1}\right)}\,,\label{eq:d_0}
\\
%
%
\bar{b}_{1}&=
\frac{
2\bar{E}_{i_I}\left(\bar{\Delta}-\bar{E}_{i_{I}}\right) 
+
\left(\gamma \bar{E}^{(-)}
- \bar{E}_{i_{I}}^2
\right)
}
{2\left(\bar{\Delta}-\bar{E}_{i_{I}+1}\right)
\left(\bar{\Delta}-\bar{E}_{i_{I}}\right)
\left(\bar{\Delta}-\bar{E}_{i_{I}-1}\right)}\,,
\label{eq:d_1}
\\
%
%
\bar{b}_{2}&=
\frac{
\left(\gamma \bar{E}^{(-)}
 - \bar{E}_{i_{I}}^2\right)
}
{2 \left(\bar{\Delta}-\bar{E}_{i_{I}+1}\right)\left(\bar{\Delta}-\bar{E}_{i_{I}}\right)\left(\bar{\Delta}-\bar{E}_{i_{I}-1}\right)}\,.
\label{eq:d_2}
\end{align}

\begin{figure}[]
\centering
\includegraphics[width=1.\linewidth]{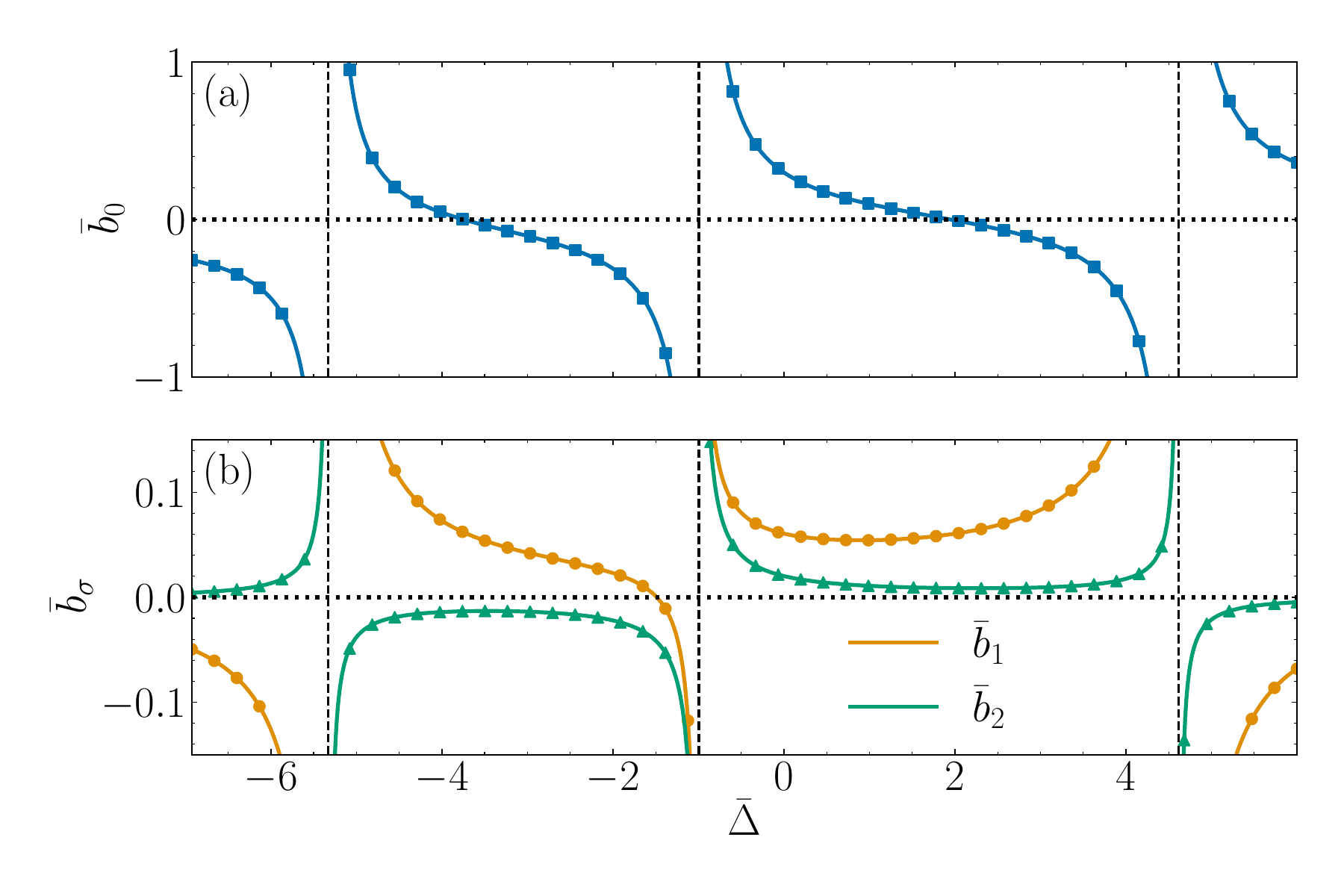}
\caption{
Dependence of the rescaled scalar $\bar{b}_0$ (a), vector $\bar{b}_1$ and tensor $\bar{b}_2$ coefficients (b) on the rescaled detuning $\bar{\Delta}$ for $i_I=9/2$ and $\gamma = 0.006$, corresponding to $^{87}$Sr atoms.
The analytical results given by~(\ref{eq:d_0}), (\ref{eq:d_1}) and (\ref{eq:d_2}) are marked by solid lines while squares, circles and triangles, respectively, represent Clebsch-Gordan numerical calculations.
}
\label{fig:fig2}
\end{figure}

In Fig.~\ref{fig:fig2} we plot the dimensionless coefficients $\bar{b}_{0,1,2}$ as a function of the relative detuning $\bar{\Delta}$ using Eqs.~(\ref{eq:d_0})-(\ref{eq:d_2}) and the conventional Clebsch-Gordan approach outlined in Appendix~\ref{app:Clebsch-Gordan}. 
There is a perfect agreement with a relative deviation due to finite computing precision below $10^{-13}$.

The analytical expressions~\eqref{eq:d_0}-\eqref{eq:d_2} show explicitly the detuning dependence of the coefficients $\bar{b}_{0,1,2}$ describing, respectively, the scalar, vector and tensor light shifts. For example, by taking the $^{87}$Sr atom with  $i_I=9/2$ and choosing detuning comparable with the HF energies, $\bar{\Delta}=\sum_{f=i_I, i_I\pm 1} \bar{E}_f/3\approx -0.57$, the dimensional coefficients are:
$\bar{b}_0 \approx 0.76$,
$\bar{b}_1 \approx 0.09$ and
$\bar{b}_2 \approx 0.05$. 
Thus the scalar term dominates for the specific value of $\bar{\Delta} $ chosen. 
Yet by changing $\bar{\Delta}$ it is possible to find a point where the scalar term $\bar{b}_0$ changes the sign and thus goes through the zero point, as one sees in Fig.~\ref{fig:fig2}. Hence one can eliminate the scalar term without cancelling the other terms.

Neglecting the quadratic hyperfine coupling ($\gamma=0$ and thus $\bar{E}_{i_{I}}=-1$), Eqs.~\eqref{eq:d_1}-\eqref{eq:d_2} shows that
the ratio of the vector and tensor light shifts goes linearly with the detuning:
\begin{equation}
 \bar{b}_1/ \bar{b}_2 =
2\bar\Delta + 3\,.
\label{b_1/b_2}
\end{equation}
Such a linear dependence holds well also after including the quadratic hyperfine coupling which is typically small ($\gamma \ll 1$). From Eq.~\eqref{b_1/b_2} it follows that the ratio $\bar{b}_1/ \bar{b}_2$ goes to zero at $\bar\Delta = -3/2$ for $\gamma=0$.

For far-off resonance radiation when detuning is much larger than the energy of the hyperfine splitting $\bar{|\Delta}| \gg 1$, the dimensionless coefficients read up to terms cubic in detuning
\begin{align}
\bar{b}_0 &\approx
\bar{\Delta}^{-1} + \frac{2}{3}\gamma \Isq 
\bar{\Delta}^{-2} +
\frac{2}{3} \Isq \left[1 + \gamma (\gamma \Isq - 1)\right] \bar{\Delta}^{-3},
\\
\bar{b}_1 &\approx
-\left(1 - \frac{\gamma}{2} \right) \bar{\Delta}^{-2} +
\frac{1}{2} \left[1 - 4 \gamma \Isq + \gamma^2 \left(3 \Isq-1\right)\right] \bar{\Delta}^{-3} ,
\\
\bar{b}_2 & \approx
-\frac{\gamma}{2} \bar{\Delta}^{-2} +
\frac{1}{2} \left[-1 +4 \gamma-\gamma^2 \left(\Isq+3\right)\right] \bar{\Delta}^{-3}.
\end{align}
In this way, for large detuning, $\bar{\Delta}\gg1$, the scalar light shift $\propto \bar{\Delta}^{-1}$ decreases linearly with the inverse detuning, whereas the vector light shift $\propto \bar{\Delta}^{-2}$ goes quadratically.
For intermediate detuning $1 \ll \bar{\Delta} \ll\gamma^{-1}$ the tensor light shift decreases as $\bar{\Delta}^{-3}$ and then cross over to an asymptotic decay $\propto \gamma/\bar{\Delta}^2$ for $\bar{\Delta} \gg\gamma^{-1}$, with $|\bar{b}_2/\bar{b}_1|\ll 1$ ensured by the weaker quadratic hyperfine coupling (typically $\gamma\ll1$).

\subsection{Effects of spontaneous emission }
\label{losses}

Let us now include the effects of the finite radiative lifetime of the excited state. This can be done by replacing the detuning $\Delta$ by $\Delta - i \hbar \Gamma$ in our treatment including the expressions for the coefficients $a_{0,1,2}$ and $b_{0,1,2}$, where $\Gamma$ is the excited state linewidth due to spontaneous emission. 
Consequently, the effective Hamiltonian acquires a non-Hermitian part representing the radiative losses.

To estimate the losses let us expand ${\rm Im}\, \bar{b}_{0,1,2}$ to first order in the linewidth $\Gamma$ assuming that the detuning is not too close to the hyperfine lines ($|\Delta - E_f| \gg \hbar \Gamma$) and neglecting the quadratic contribution to the hyperfine coupling ($\gamma = 0$):
\begin{equation}
{\rm Im}\, \bar{b}_0 \approx 
\frac{
3 (\bar{\Delta} + 1)^4 +
2 (\bar{\Delta} + 1)\Isq + 
\Iqu}
{3 \left(\bar{\Delta}-\bar{E}_{i_{I}+1}\right)^2\left(\bar{\Delta}-\bar{E}_{i_{I}}\right)^2\left(\bar{\Delta}-\bar{E}_{i_{I}-1}\right)^2}
\bar{\Gamma}\,,
\label{eq:im-b0-expansion}
\end{equation}
\begin{equation}
{\rm Im}\, \bar{b}_1 \approx 
-\frac{
4\bar{\Delta}^3 +
13\bar{\Delta}^2 +
12\bar{\Delta}
-\Isq +3
}
{2 \left(\bar{\Delta}-\bar{E}_{i_{I}+1}\right)^2\left(\bar{\Delta}-\bar{E}_{i_{I}}\right)^2\left(\bar{\Delta}-\bar{E}_{i_{I}-1}\right)^2}
\bar{\Gamma}\,,
\label{eq:im-b1-expansion}
\end{equation}
\begin{equation}
{\rm Im}\, \bar{b}_2 \approx
-\frac {3\bar{\Delta}^2 + 4\bar{\Delta} - \Isq + 1}
{2 \left(\bar{\Delta}-\bar{E}_{i_{I}+1}\right)^2\left(\bar{\Delta}-\bar{E}_{i_{I}}\right)^2\left(\bar{\Delta}-\bar{E}_{i_{I}-1}\right)^2}
\bar{\Gamma}\,,
\label{eq:im-b2-expansion}
\end{equation}
with $\bar{\Gamma} = \Gamma / A_{\rm HF}$.

Although these formulas do not hold when the detuning $|\Delta - E_f|$ becomes comparable to $\hbar \Gamma$ or is smaller, such a situation is not interesting because of a significant increase of losses compared to the light shifts. On the other hand, the intermediate ($\hbar \Gamma \ll |\Delta - E_f| \ll \hbar A_{\rm HF}$) and far detuned regimes are well described by Eqs.~\eqref{eq:im-b0-expansion}-\eqref{eq:im-b2-expansion}. For the far detuned radiation, $|\Delta - E_f| \gg \hbar A_{\rm HF}$, the losses go as  ${\rm Im}\,\bar{b}_0 \propto \bar{\Delta}^{-2}$, ${\rm Im}\,\bar{b}_1 \propto \bar{\Delta}^{-3}$ and ${\rm Im}\,\bar{b}_2 \propto \bar{\Delta}^{-4}$, so the dominant loss term ${\rm Im}\,\bar{b}_0$ and the vector light shift coefficient ${\rm Re}\ \bar{b}_1$ both go as $1/\Delta^2$.
Hence, in the far detuned regime the ratio ${\rm Im}\bar{b}_0/ {\rm Re} \bar{b}_1 $  becomes constant  and is determined by the ratio of the linewidth $\Gamma$ and the hyperfine structure constant $A_{\rm HF}$: 
\begin{equation}
{
 |{\rm Im}\bar{b}_0 / {\rm Re}\ \bar{b}_1| 
  \approx \Gamma / |A_{{\rm HF}}| \ll 1
\,.
\label{ratio}
}
\end{equation}
For example, for the 
$^1S_0 \rightarrow ^3P_1 $
line of the species: spin-7/2 $^{43}$Ca; spin-9/2 $^{87}$Sr; spin-1/2 $^{171}$Yb; spin-5/2 $^{173}$Yb, this ratio is, respectively, $ \Gamma / |A_{{\rm HF}}|= 2.\,10^{-6}; 3.\,10^{-5}; 5.\,10^{-5}; 2.\,10^{-4}$, showing that the intercombination line of alkaline-earth-like atoms is very favorable to engineer nuclear-spin-sensitive vector or tensor light shifts.

\begin{figure}[]
\includegraphics[width=1.\linewidth]
{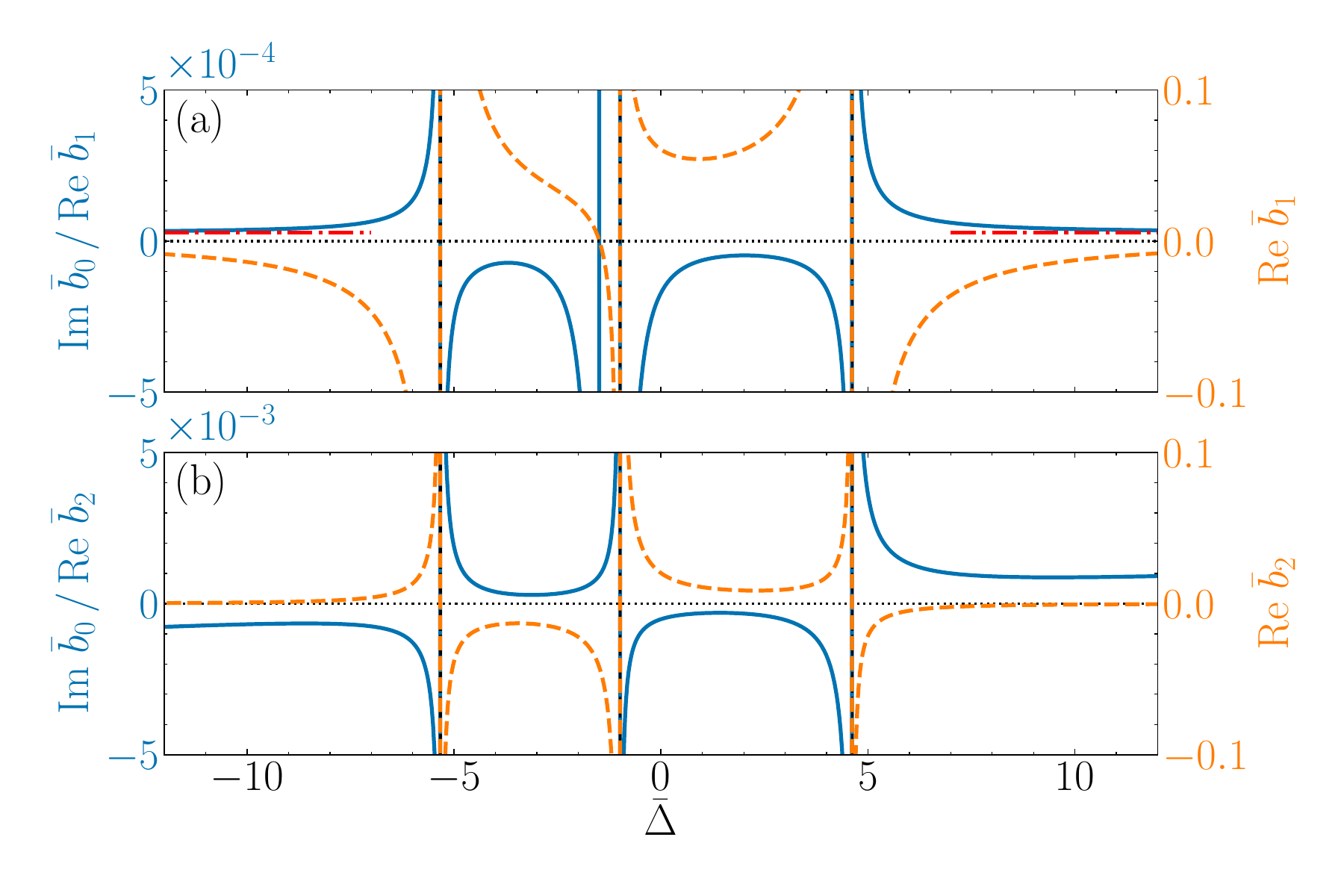}
\caption{
Dependence of the loss ratio ${\rm Im}\ \bar{b}_0/{\rm Re}\ \bar{b}_{1,2}$ (blue solid lines, left vertical axis) and the dimensionless vector (a) and tensor (b) polarizabilities ${\rm Re} \, \bar{b}_{1,2}$ (dashed yellow lines, the right vertical axis) on the dimensionless detuning $\bar\Delta$ for $\Gamma / |A_{{\rm HF}}|=3\times10^{-5}$.
Red dash-dotted horizontal lines represent the limiting value  of the ratio ${\rm Im}\ \bar{b}_0/{\rm Re}\ \bar{b}_{1}$ for large detuning given by  Eq.~\eqref{ratio}.}
\label{fig:fig-vector-ratio-SE-version}
\end{figure}

Yet in the far-detuned regime, the vector and tensor light shits decrease considerably. To get their larger values, one needs to go closer to the resonances or even between the resonance lines. In that case the relative contribution of the losses can still have reasonably small (of the order of $10^{-4}$) with a considerably increase of the vector light polarizability, 
as one can see in Fig.~\ref{fig:fig-vector-ratio-SE-version}.

\section{Specific configurations}
\label{sec:three}

The analytical form of the effective Hamiltonian~(\ref{eq:generalDij}) of the ground state manifold, can be presented as 
\begin{equation}
\hat{H}_{{\rm eff}}=\hat{H}_{{\rm eff}(0)}+\hat{H}_{{\rm eff}(1)}+\hat{H}_{{\rm eff}(2)}\,,\label{eq:H_eff-symmetric}
\end{equation}
where
\begin{equation}
\hat{H}_{{\rm eff}(0)}
\coloneqq
\frac{b_0}{4}\left(\boldsymbol{\mathcal{E}}^{*}\cdot\boldsymbol{\mathcal{E}}\right)=\frac{b_0}{4}\left|\boldsymbol{\mathcal{E}}\right|^{2}\,,\label{eq:H_eff-0}
\end{equation}
is the scalar component,
\begin{equation}
\hat{H}_{{\rm eff}(1)}\coloneqq
\frac{i b_1}{4\hbar}
\textbf{\ensuremath{\mathbf{I}}}\cdot\left(\boldsymbol{\mathcal{E}}^{*}\times\boldsymbol{\mathcal{E}}\right)
=
-\frac{b_{1}}{2\hbar}
\textbf{\ensuremath{\mathbf{I}}}\cdot\left(\boldsymbol{\mathcal{E}}^{\prime}\times\boldsymbol{\mathcal{E}}^{\prime\prime}\right)\,,\label{eq:H_eff-1}
\end{equation}
is the vector component, and
\begin{align}
\hat{H}_{{\rm eff}(2)}\coloneqq
&
\frac{b_{2}}{4\hbar^2}
\left[ 
\left(\boldsymbol{\mathcal{E}}^{*}\cdot\textbf{\ensuremath{\mathbf{I}}}\right)\left(\boldsymbol{\mathcal{E}}\cdot\textbf{\ensuremath{\mathbf{I}}}\right)
-\frac{1}{3}\left|\boldsymbol{\mathcal{E}}\right|^{2}{\mathbf{I}}^2 + {\rm h.c.}
\right] \nonumber \\
= 
& \frac{b_{2}}{2\hbar^{2}}
\left[
\left(\boldsymbol{\mathcal{E}}^{\prime}\cdot\textbf{\ensuremath{\mathbf{I}}}\right)^{2}+\left(\boldsymbol{\mathcal{E}}^{\prime\prime}\cdot\textbf{\ensuremath{\mathbf{I}}}\right)^{2}-\frac{1}{3}\left|\boldsymbol{\mathcal{E}}\right|^{2}{\mathbf{I}}^2
\right] 
\,,\label{eq:H_eff-2}
\end{align}
is the tensor component.
In the above expressions, we used the real and imaginary parts of the complex light field $\boldsymbol{\mathcal{E}}=\boldsymbol{\mathcal{E}}^{\prime}+i\boldsymbol{\mathcal{E}}^{\prime\prime}$.
This entirely analytical form 
allows one to explore its form to design desired atom-light coupling. Below we discuss three example configurations including a single beam and a pair of counter-propagating beams and a pair of perpendicular beams. The last configuration is demonstrated to generate spin-orbit coupling for the nuclear spin. Finally in Subsection~\ref{sec:MARTIN} we discuss a way of eliminating the tensor light shifts by combining light fields with highly different frequencies.

\subsection{Single beam}
\label{sebsec:Single beam}

Let us consider first a situation where the system is subjected to a single linearly polarized laser beam
\begin{equation}
\boldsymbol{\mathcal{E}}=\mathcal{E}e^{ikz}\boldsymbol{\mathrm{e}}_{z}\,.
\label{eq:E-single_beam-linear}
\end{equation}
where the amplitude $\mathcal{E}$ is assumed to be real and no extra global phase is added. This can be done by properly choosing the origin of spatial coordinates in all the examples considered here in Secs.~\ref{sebsec:Single beam}-\ref{sec:Perpendicular beams}.

For the single beam the vector light shift (\ref{eq:H_eff-1}) is zero and effective Hamiltonian contains only the position-independent scalar and tensor components
\begin{equation}
\hat{H}_{{\rm eff}}=\frac{\mathcal{E}^{2}}{4}
\left[b_{0} 
\mathbb{1} 
+
\frac{2b_{2}}{\hbar^{2}}
\left(
{\bf I}_{z}^{2}-\frac{{\bf I}^{2}}{3}
\right) 
\right]\,,\label{eq:H_eff-single_beam-linear}
\end{equation}
where $\mathbb{1}$ is an identity operator.
This shifts the ground-state energies
according to $\hat{H}_{{\rm eff}} |g\rangle |i_I, m_I\rangle = E^{(i_I)}_{m_I}|g\rangle |i_I, m_I\rangle$ with eigenvalues having a scalar term (independent of $m_I$) and tensor shifts (quadratic in $m_I$), namely
$E^{(i_I)}_{m_I}=\mathcal{E}^{2}/4
\left[b_{0} 
+
2b_{2}
\left(
m_{I}^{2}-\Isq/3
\right) 
\right]$.

In the case of a circularly polarized laser beam,
\begin{equation}
\boldsymbol{\mathcal{E}}=\mathcal{E}e^{ikz}\boldsymbol{e}_{\pm}\,,\quad\mathrm{with}\quad\boldsymbol{e}_{\pm}=\frac{1}{\sqrt{2}}\left(\boldsymbol{\mathrm{e}}_{x}\pm i\boldsymbol{\mathrm{e}}_{y}\right)\,,
\label{eq:E-single_beam-circular}
\end{equation}
the effective Hamiltonian contains additionally the (position-independent) vector components $\propto {\bf I}_{z}$:
\begin{equation}
\hat{H}_{{\rm eff}}=\frac{\mathcal{E}^{2}}{4}\left[b_{0}\mathbb{1} 
\mp
\frac{b_{1}}{\hbar}{\bf I}_{z}+\frac{b_{2}}{\hbar^{2}}\left( 
\frac{1}{3}{\bf I}^{2}-{\bf I}_{z}^{2}
\right)
\right]\,.\label{eq:H_eff-single_beam-circular}
\end{equation}
In this configuration, eigenenergies of the ground states manifold $|g\rangle |i_I, m_I\rangle$ contain additional position-independent linear (vector) shifts,
$E^{(i_I)}_{m_I}=\mathcal{E}^{2}/4
\left[b_{0} 
\mp b_1 m_I
-
b_{2}
\left(
m_{I}^{2}-\Isq/3
\right) 
\right]$.

\subsection{Two cross polarized counter-propagating laser beams}
\label{sebsec:twocounterpropagating}

Consider next the effect of two cross-polarized counter-propagating Raman beams where
\begin{equation}
\boldsymbol{\mathcal{E}}=
\frac{\mathcal{E}^{2}}{\sqrt{2}}
\left( e^{ikz}\boldsymbol{\mathrm{e}}_{x}+e^{-ikz}\boldsymbol{\mathrm{e}}_{y}\right) \,.
\label{eq:E-counterpropagating}
\end{equation}

In that case, the scalar light shift is uniform
\begin{equation}
\hat{H}_{{\rm eff}(0)}
=\frac{\mathcal{E}^{2} }{4}b_{0}\mathbb{1}\,
\label{eq:H_eff-0-counterpropagating}
\end{equation}
and there are spatially periodic vector and tensor components 
\begin{equation}
\hat{H}_{{\rm eff}(1)}=
\frac{\mathcal{E}^{2} }{4}\frac{b_{1}}{\hbar}
\sin\left(2kz\right){\bf I}_{z}\,.\label{eq:H_eff-1-counterpropagating}
\end{equation}
and
\begin{equation}
\hat{H}_{{\rm eff}(2)}
=\frac{1}{2}\frac{b_{2}}{\hbar^{2}}\mathcal{E}^{2}
\left(
\cos^{2}\left(kz\right){\bf I}_{\tilde{y}}^{2}+\sin^{2}\left(kz\right){\bf I}_{\tilde{x}}^{2}-\frac{1}{3}{\bf I}^{2}
\right) \,,
\label{eq:H_eff-2-counterpropagating}
\end{equation}
with
\begin{equation}
{\bf I}_{\tilde{x}}\coloneqq\frac{{\bf I}_{x}-{\bf I}_{y}}{\sqrt{2}} \quad \mathrm{and} \quad {\bf I}_{\tilde{y}}\coloneqq\frac{{\bf I}_{x}+{\bf I}_{y}}{\sqrt{2}}\,.
\end{equation}
The latter $\hat{H}_{{\rm eff}(2)}$ can be also represented in terms of the original nuclear spin operators ${\bf I}_{x,y,z}$: 
\begin{equation}
\hat{H}_{{\rm eff}(2)}
=\frac{1}{2}\frac{b_{2}}{\hbar^{2}}\mathcal{E}^{2}
\left(
\frac{1}{6}{\bf I}^2 - \frac{1}{2}{\bf I}_z^2 + \frac{\cos\left(2kz\right)}{2} \left\{{\bf I}_x,{\bf I}_y\right\}
\right) \,,
\label{eq:labframe}
\end{equation}
where anti-commutators can be rephrased by using the commutation relations: $\{{\bf I}_x,{\bf I}_y\}=[{\bf I}_y,{\bf I}_x]+2{\bf I}_x{\bf I}_y=-i\hbar {\bf I}_z+2{\bf I}_x{\bf I}_y$.

Therefore in addition to uniform position-independent light shifts, 
in this configuration involving counterpropagating light beams
there is also a state-dependent periodic lattice potential due to the scalar and vector terms $\hat{H}_{{\rm eff}(1)}$ and $\hat{H}_{{\rm eff}(2)}$, Eqs.~\eqref{eq:H_eff-1-counterpropagating}-\eqref{eq:labframe}.


\subsection{Two perpendicular light beams for spin orbit coupling} \label{sec:Perpendicular beams}

In the typical situations presented in the previous subsections, the effective Hamiltonian contains both vector and tensor shifts.
Here we will consider a scheme in which these light shifts describe the spin-orbit coupling (SOC) of the NIST type \cite{Goldman:2014RPP, Lin2011} supplied with an extra quadratic term (tensor light shift). For additional flexibility, we will allow a small frequency difference $\delta \omega$ with  $|\delta \omega/A_{\rm HF}|\ll1$ between two beams enabling us to remove or modify the position-independent linear shift in the effective Hamiltonian.

The scheme involves two light beams of equal amplitudes propagating at a right angle. 
One of them is propagating along the $z$ axis and is circularly $\boldsymbol{e}_{+}$ polarized. The second beam is propagating along the $y$ axis with a linear polarization $\boldsymbol{e}_{z}$ and a frequency difference $\delta \omega$ with respect the the first beam, so that
\begin{equation}
\boldsymbol{\mathcal{E}}=\frac{\mathcal{E}}{\sqrt{2}}\left(\boldsymbol{e}_{+}e^{ikz}+\boldsymbol{e}_{z}e^{i(ky-\delta\omega t)}\right)\,.\label{eq:E-90degrees}
\end{equation}
The scalar light shift is then uniform,
$
\hat{H}_{{\rm eff}(0)}=\frac{1}{4}b_{0}\mathcal{E}^{2}
$, and thus will be omitted by redefining the detuning energy $\Delta$.
The vector and tensor components are
\begin{align}
\hat{H}_{{\rm eff}(1)}&=-
\frac{\mathcal{E}^{2}}{8}\frac{b_{1}}{\hbar}\left( {\bf I}_{z}-\sqrt{2}
{\bf I}_{xy}\left(s\right)\right),
\label{eq:H_eff_1-90-degrees}\\
\hat{H}_{{\rm eff}(2)}&=
-\frac{\mathcal{E}^{2}}{4} 
\frac{b_{2}}{\hbar^{2}}
\left(
\frac{{\bf I}^{2}}{6}
-
\frac{{\bf I}_{z}^{2}}{2}
-
\frac{\{{\bf I}_{xy} \left(s\right),{\bf I}_z\}}{\sqrt{2}}
\right)
\,,\label{eq:H_eff-2-90-degrees-alt}
\end{align}
where
\begin{equation}
{\bf I}_{xy}\left(s\right) \coloneqq {\bf I}_x \cos\left(s\right) + {\bf I}_y \sin\left(s\right)\,
\label{I_xy}
\end{equation}
and $s = ky-kz-\delta\omega t$. 
Alternatively the operator ${\bf I}_{xy}\left(s\right)$ can be cast in terms of the spin raising and lowering operator ${\bf I}_{\pm}={\bf I}_{x}\pm i{\bf I}_{y}$ as
\begin{equation}
{\bf I}_{xy} \left(s\right) =
{\bf I}_{+}e^{is}+{\bf I}_{-}e^{-is} \,.\label{I_xy_alternative}
\end{equation}

The time dependence of linear and quadratic shifts can be eliminated by a unitary transformation describing the time-dependent spin rotation around the $z$ axis
\begin{equation}
\hat U = \exp\left(
-i {\bf I}_z \delta\omega t / \hbar
\right)\,.
\end{equation}
Consequently ${\bf I}_{xy}(s)$ entering $\hat{H}_{{\rm eff}(1)}$ and $\hat{H}_{{\rm eff}(2)}$ transforms into the time-independent oprator
$\hat U^\dagger {\bf I}_{xy}(s) \hat U = {\bf I}_{xy}(ky-kz)$, and the light shifts take the form:
\begin{equation}
\hat{H}_\mathrm{eff(1)} = -\frac{\mathcal{E}^{2}}{8}\frac{b_{1}}{\hbar}\left[ 
{\bf I}_{z}
-\sqrt{2}
{\bf I}_{xy}(ky-kz)\right] + \delta\omega {\bf I}_z \,
\label{eq:H_eff_trans_1}
\end{equation}
and
\begin{equation}
\hat{H}_\mathrm{eff(2)} = 
\frac{\mathcal{E}^{2}}{4} 
\frac{b_{2}}{\hbar^{2}} 
\left[
-\frac{{\bf I}^{2}}{6} +
\frac{{\bf I}_{z}^{2}}{2}
+\frac{\{{\bf I}_{xy}(ky-kz),{\bf I}_z\}}{\sqrt{2}}
\right]\,,
\label{eq:H_eff_trans_2}
\end{equation}
where an extra linear Zeeman shift term $\delta\omega {\bf I}_z$ appears in $\hat{H}_{{\rm eff}(1)}$ due to the time-dependence of the unitary transformation $\hat U$. In particular, by choosing $\delta\omega = b_1\mathcal{E}^{2}/8\hbar$, the linear shift $\propto {\bf I}_z$ is eliminated in Eq.~\eqref{eq:H_eff_trans_1}, giving
\begin{equation}
\hat{H}_\mathrm{eff(1)} =  \frac{\mathcal{E}^{2}}{4\sqrt{2}}\frac{b_{1}}{\hbar} {\bf I}_{xy}(ky-kz) \,.
\label{eq:H_eff_trans_1-without-quadratic}
\end{equation}

The operator ${\bf I}_{xy}(ky-kz)$ defined by Eq.~\eqref{I_xy} and featured in $\hat{H}_{{\rm eff}(1)}$ and $\hat{H}_{{\rm eff}(2)}$ represents the spin rotating in the $xy$ plane when the atomic position changes in the $y-z$ direction.
This provides a linear SOC of the NIST type \cite{Goldman:2014RPP, Lin2011} in the vector component of the effective Hamiltonian $\hat{H}_{{\rm eff}(1)}$ given by Eqs.~\eqref{eq:H_eff_trans_1} or \eqref{eq:H_eff_trans_1-without-quadratic}, and a more complex quadratic SOC in the tensor component $\hat{H}_{{\rm eff}(2)}$ given by Eq.~\eqref{eq:H_eff_trans_2}. Note that the position dependence of both linear and quadratic terms can be eliminated by an additional unitary operation $\hat U _1= \exp\left[
-i (kz-ky) {\bf I}_z / \hbar
\right]$ transforming ${\bf I}_{xy}(ky-kz)$ to the position-independent spin operator ${\bf I}_{x}$. As in the case of spin-1/2 \cite{Goldman:2014RPP,Lin2011}, the SOC is then represented by the spin-dependent momentum shift when the atomic kinetic energy is included.

\subsection{Combining significantly different frequencies for eliminating tensor light shifts}
\label{sec:MARTIN}

As exemplified above, realising purely linear light shifts, in particular the linear SOC, can be complicated for atoms with a large spin, as the tensor terms are present in typical configurations. It has been proposed to tune these out by stroboscopically alternating different polarisation configurations at a rate exceeding the atomic response rate~\cite{Mamaev2022}. As an alternative relying on continuous illumination, we propose to use a bichromatic light with two widely different frequencies, ensuring that the resulting vector light shifts are of the same sign. Yet, perfectly opposing tensor terms cancel each other. 

In this way, let us consider two laser fields $\alpha$ and $\beta$ with the same configurations, but significantly different detunings $\Delta_{\alpha}$ and $\Delta_{\beta}$.
Specifically the frequency difference $\Delta_{\alpha} - \Delta_{\beta}$ should be large compared to the characteristic frequencies of the vector and tensor light shifts, so that the fast oscillating cross terms of the effective Hamiltonian average out.  
The resulting effective Hamiltonian is thus additive, namely is given by the sum of the effective Hamiltonians due to the separate fields $\alpha$ and~$\beta$,
\begin{equation}
    \hat{H}_{\rm eff}=\hat{H}_{{\rm eff},\alpha} + \hat{H}_{{\rm eff}, \beta}.
\label{H_eff-bichromatic}
\end{equation}
In all situations considered in Secs.~\ref{sebsec:Single beam}-\ref{sec:Perpendicular beams}, the linear and quadratic light shifts are characterised by the factors $\mathcal{E}^2 b_1 \equiv \mathcal{E}^2 b_1(\Delta)$ and $\mathcal{E}^2 b_2 \equiv \mathcal{E}^2 b_2(\Delta)$. For the bichromatic field these factors are to be replaced by $\mathcal{E}_\alpha^2 b_1(\Delta_\alpha)+ \mathcal{E}_\beta^2 b_1(\Delta_\beta)$ and $\mathcal{E}_\alpha^2 b_2(\Delta_\alpha)+ \mathcal{E}_\beta^2 b_2(\Delta_\beta)$, respectively, where the normalization condition $\mathcal{E}_\alpha^2 + \mathcal{E}_\beta^2 = 1$ applies to the amplitudes of the electric field. 
The quadratic light shift can be eliminated by properly choosing detunings and amplitudes of the electric fields such that $\mathcal{E}_\alpha^2 {\rm Re} [b_2(\Delta_\alpha)]+ \mathcal{E}_\beta^2 {\rm Re} [ b_2(\Delta_\beta) ] = 0$.

\begin{figure}
{
\centering
\includegraphics[width=1.\linewidth]
{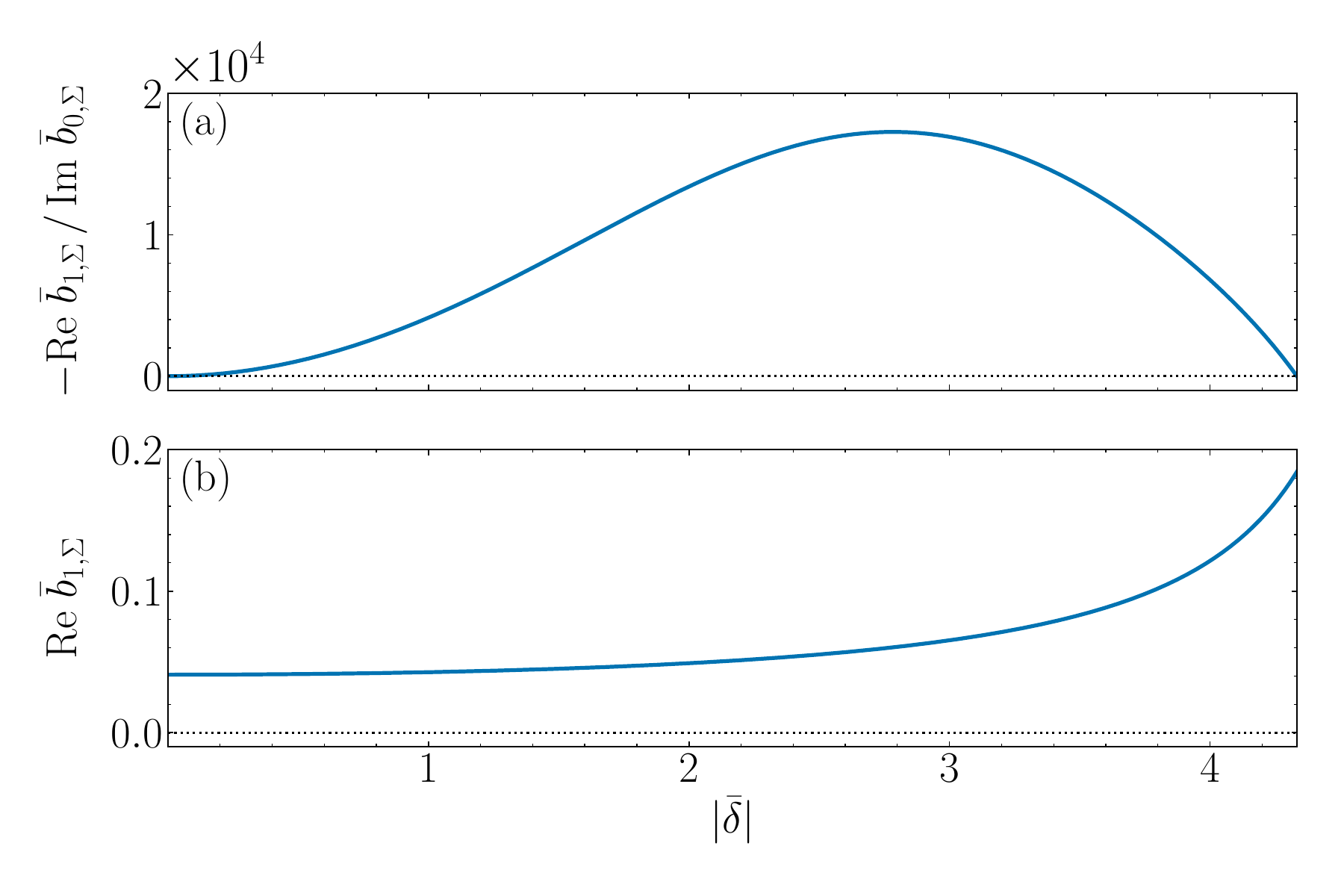}
\caption{
The ratio ${\rm Re}\,\bar{b}_{1,\Sigma} / {\rm Im}\,\bar{b}_{0,\Sigma}$ (a, upper panel)  
and
${\rm Re}\,\bar{b}_{1,\Sigma}$
(b, lower panel) vs the dimensionless detuning imbalance $\bar{\delta}=\delta / (\hbar A_{\rm HF})$,
where
$\bar{b}_{i,{\rm \Sigma}} = \mathcal{E}_\alpha^2 \bar{b}_i(\Delta_\alpha)+ \mathcal{E}_\beta^2 \bar{b}_i(\Delta_\beta)$ (with i=0,1,2) are the dimensionless bichromatic coefficients. 
Here the normalization condition $\mathcal{E}_\alpha^2 + \mathcal{E}_\beta^2 = 1$ is used and the calculations were carried out for $\Gamma /|A_{\rm HF}| \approx 3\times 10^{-5}$.
The amplitudes of bichromatic light beams were chosen such that the condition $\mathcal{E}_\alpha^2 {\rm Re}\,\bar{b}_2(\Delta_\alpha)+ \mathcal{E}_\beta^2 {\rm Re}\,\bar{b}_2(\Delta_\beta)=0$ holds, and the tensor light shift is eliminated. 
 This requirement can be always satisfied for values of $\bar{\delta}$ shown in this plot, but not for arbitrary $\bar{\delta}$, as beyond this region signs of ${\rm Re}\, \bar{b}_2(\Delta_\alpha)$ and ${\rm Re}\, \bar{b}_2(\Delta_\beta)$ may become the same.
}
\label{fig:bichro-b1-ratio-uSE}
}
\end{figure}

The analytical expressions for the vector and tensor coefficients $b_{1}$ and $b_2$ make it easy to determine suitable configurations. In particular, we note that $b_2$ changes sign when crossing the central hyperfine line at $\Delta = E_{i_I}$, while $b_1$  does not (apart from a tiny spectral window), as one can see in Fig.~\ref{fig:fig2} and~\ref{fig:fig-vector-ratio-SE-version}. Thus one can eliminate the quadratic light shift by taking the detunnings $\Delta_\alpha$ and $\Delta_\beta$ to be on different sides of the central hyperfine line. For example, one can choose
detunings to be symmetric with respect to the central hyperfine line: $\Delta_\alpha = E_{i_I} + \delta$ and $\Delta_\beta = E_{i_I} - \delta$ \footnote{The frequency difference $\delta$ in the bichromatic scheme should not be confused with a much smaller frequency difference $\delta \omega$ included in the scheme considered in Sec~\ref{sec:Perpendicular beams} to have an adjustable linear Zeeman shift.}.

As shown in Fig.\,\ref{fig:bichro-b1-ratio-uSE}, for $i_I = 9/2$ corresponding to $^{87}$Sr, at $\delta_{\rm optimal} \simeq 3 |A_{\rm HF}|$ there is a local optimal ratio between vector shift and loss rate due to spontaneous emission, ${\rm Re}\,\bar{b}_{1,\Sigma} / {\rm Im}\,\bar{b}_{0,\Sigma}$, is $10^4 $, where
$\bar{b}_{i,{\rm \Sigma}} = \mathcal{E}_\alpha^2 \bar{b}_i(\Delta_\alpha)+ \mathcal{E}_\beta^2 \bar{b}_i(\Delta_\beta)$ (with i=0,1,2) are the dimensionless bichromatic coefficients.

Using two light fields even further detuned, away from the hyperfine structure,  one can also have the opposite signs of $b_2$ and same signs of $b_1$ needed for the cancelation of the tensor light shifts without cancelling the vector ones. There are however two drawbacks:~i) vector shifts smaller in absolute value,~ii) a complication in the implementation of spin-dependent lattices or vector spin-orbit coupling. Indeed, the schemes considered in Sections~\ref{sebsec:twocounterpropagating} and \ref{sec:Perpendicular beams} involve a local phase related to optical path differences between two beams. When superimposing two spin-dependent lattices created with light beams at differing wavelengths, the slight difference $k= k_\alpha - k_\beta$  in wavevectors $k_\alpha$ and $k_\beta$ might matter. 
These difficulties can be overcome by working at the relatively small detuning within the hyperfine structure, as discussed above ($\delta_{\rm optimal} \simeq 3 |A_{\rm HF}|$ for $i_I = 9/2$).
In that case, one can ensure that the two polarization lattices are in phase over the lengthscale of the atomic cloud.
For example, in the counterpropagating beam configuration, where the optical path difference $2 k (z-z_0)$ is typically controlled by the position $z_0$ of a mirror, this rephasing occurs every $z-z_0 = n \times \pi c / 2 \delta $ with $n \in \mathbb{N}$. For $^{87}$Sr, one has $\pi c / 2 \delta_{\rm optimal} \simeq 10\,$cm. This is a macroscopic distance, so it is possible to tune the optical path length accurately enough and thus realize a good cancellation of the tensor light shift.

\section{Summary and discussions}
\label{sec:four}

We have analyzed the effective Hamiltonian describing the scalar, vector, and tensor light shifts of nuclear spin states in alkaline earth atoms. Our approach, rooted in the Dyson identity, circumvents the need for explicit scrutiny of transitions within the excited state manifold's specific hyperfine states. Instead, we have derived simple analytical expressions for the coefficients defining the scalar, vector and tensor light shift in the effective Hamiltonian.
Our work on one-body Hamiltonians can be a stepping stone to design experiments where ad-hoc one-body terms (such as SOC) combined with inter-atomic interactions lead to emerging phenomena (quantum phases, spintronic applications). 

We have explored typical examples of light configurations capable of inducing light shifts for the nuclear spin states in the ground electronic state manifold. Additionally, we have discerned configurations governing SOC that encompass both linear and quadratic shifts, alongside those characterized by purely linear SOC effects.

The breadth of our analytical findings is broad-reaching and can be applied to other alkaline earth (or alkaline earth-like) atoms with analogous structures. We anticipate that the straightforward analytical expressions for scalar, vector, and tensor terms we have provided will prove helpful in guiding future experimental endeavours and fostering a more profound theoretical comprehension of systems subjected to off-resonance light coupling.

\section*{ACKNOWLEDGMENTS}
We gratefully acknowledge discussions with Bruno Laburthe-Tolra.
This work was supported by DAINA project of the Polish National Science Center UMO-2020/38/L/ST2/00375 and Lithuanian Research Council S-LL-21-3.
E.W. acknowledges support of the Polish National
Science Centre project UMO-2016/22/E/ST3/00045.

\appendix

\begin{widetext}

\section{Dyson equation/identity}\label{app:Dyson}

Let us consider a Hamiltonian represented in terms of the zero-order Hamiltonian $H_0$ and the interaction operator $V$ as: 
$H=H_0 + V$. We define the full Green operator
$G=(\Delta- H)^{-1}$ and the zero-order Green operator $G_0=(\Delta- H_0)^{-1}$, where $\Delta$ is some parameter in energy units; in the present situation it is the detuning. The Green operator obeys the Dyson equation (identity)
\begin{equation}
    G = G_0 + G_0 V G.
\label{eq:Dyson}
\end{equation}
The validity of the Dyson identity can be proved by rearranging the Green operator in the following way:
\begin{equation}
    G = G_0 (\Delta- H_0)G = G_0 (\Delta- H + V)G = G_0  +  G_0 V G.
\label{eq:Dyson-proof-final}
\end{equation}
The Dyson equation \eqref{eq:Dyson} is exact and valid for any $H_0$ and $V$, no approximation or assumption has been made. Note also that in this work we take $H_0=0$ and $V=H_\mathrm{HF}$. In that case, Eq.~\eqref{eq:Dyson} reduces to Dyson equation presented in the main text in Eq.~\eqref{eq:Dyson-identity-for-G}.

\section{Derivation of the the equation for $\mathbf{K}$ and its solution}
\label{sec:Append:Effective-Hamiltonian-including-quadratic-term}

\subsection{Equation for $\mathbf{K}$}
\label{subsec:Append:Equation for K}

The vector operator $\mathbf{K}$ with the Cartesian components $K_i$ is defined as 
\begin{equation}
K_{s}=D_{s,q}\mathcal{E}_{q}\,,\label{eq:F_i_app}
\end{equation}
where $D_{s,q}$ is a tensor describing the second order light induced interaction for the ground state atoms:
\begin{equation}
D_{s,q} =P_{g}d_{s} P_e G P_e  d_{q}P_{g}\,.
\label{eq:D_sq_def-app}
\end{equation}
Applying the Dyson identity for $G$ given by Eq.~\eqref{eq:Dyson-identity-for-G} or \eqref{eq:Dyson}, one gets the following relation for $D_{s,q}$:
\begin{align}
\label{eq:D_ij-equation-0:app}
D_{s,q} & =\frac{\left|d_{ge}^{2}\right|}{\Delta}
P_{g}
\delta_{sq}
+
P_{g}d_{s}P_e H_{\rm HF} G P_{e}d_{q}P_{g}\,.
\end{align}
Combining it with the expression \eqref{eq:F_i_app} for $\mathbf{K}$, one has:
\begin{align}
\mathbf{K} & =\frac{1}{\Delta}P_{g}\left|d_{ge}^{2}\right|\mathcal{\boldsymbol{\mathcal{E}}}
+\frac{A_{{\rm HF}}}{\hbar\Delta}P_{g}\mathbf{d} P_e \left(\mathbf{I}\cdot\mathbf{J}\right)\left(1+\frac{\gamma}{\hbar^{2}}\mathbf{I}\cdot\mathbf{J}\right)GP_{e}\left(\mathbf{d}\cdot\boldsymbol{\mathcal{E}}\right)P_{g}\,.\label{eq:K-equation-App}
\end{align}
Our aim is get a closed equation [Eq.~\eqref{eq:K-vector-equation-2-App}] for $\mathbf{K}$ containing only the spin operators $\mathbf{I}$ and the electric field amplitude $\mathcal{\boldsymbol{\mathcal{E}}}$.
For this, we will carry out the following steps.
Since $\mathbf{J}$ commutes with $P_e$ and gives zero when acting on the ground state manifold ($P_{g}{\bf J}=0$), one has
 \begin{equation}
P_{g}d_{i}P_e\left(\mathbf{J}\cdot\mathbf{I}\right)=P_{g}d_{i}\left(\mathbf{J}\cdot\mathbf{I}\right)P_e=P_{g}\left[d_{i},\left(\mathbf{J}\cdot\mathbf{I}\right)\right]P_e=-i\hbar\epsilon_{lik}P_{g}d_{k}P_e I_{l},
\label{eq:replacement}
\end{equation}
where the use has been made of the commutator relations \cite{Landau:1987}:
\begin{equation}
    [J_i, d_j] = [L_i , d_j]=i \hbar \epsilon_{ijk}d_k,
    \label{Commutator}
\end{equation}
Equation \eqref{eq:replacement} can be represented in a vector form as:
\begin{equation}
P_{g}\mathbf{d} P_e \left(\mathbf{J}\cdot\mathbf{I}\right)=i\hbar P_{g}\mathbf{I}\times\mathbf{d}P_e\,.
\label{eq:replacement-1}
\end{equation}
With this Eq.~\eqref{eq:K-equation-App} reduces to
\begin{align}
\mathbf{K} =\frac{1}{\Delta}P_{g}\left|d_{ge}^{2}\right|\mathcal{\boldsymbol{\mathcal{E}}}
+i\frac{A_{{\rm HF}}}{\Delta}P_{g}\left[\mathbf{I}\times\mathbf{d}P_e+\frac{\gamma}{\hbar^{2}}\left(\mathbf{I}\times\mathbf{d}\right)P_e\left(\mathbf{I}\cdot\mathbf{J}\right)\right]GP_{e}\left(\mathbf{d}\cdot\boldsymbol{\mathcal{E}}\right)P_{g}\,.\label{eq:K-equation-App-1}
\end{align}
Using the commutation relation \eqref{Commutator}, one has for the Cartesian components of the operator $P_g\left(\mathbf{I}\times\mathbf{d}\right)P_e\left(\mathbf{I}\cdot\mathbf{J}\right)$:
\begin{equation}
P_{g}\left(\mathbf{I}\times\mathbf{d}\right)_{i}P_e\left(\mathbf{I}\cdot\mathbf{J}\right)=\epsilon_{ilk}I_{l}P_gd_{k}P_e\left(\mathbf{I}\cdot\mathbf{J}\right)=i\hbar\epsilon_{ilk}\epsilon_{kpq}I_{l}P_g d_{q}P_eI_{p}\,.
\end{equation}
Since $\epsilon_{ilk}\epsilon_{kpq}=\epsilon_{ilk}\epsilon_{pqk}=\delta_{ip}\delta_{lq}-\delta_{iq}\delta_{lp}$, then
\begin{equation}
P_{g}\left(\mathbf{I}\times\mathbf{d}\right)_{i}P_e\left(\mathbf{I}\cdot\mathbf{J}\right)=i\hbar P_g\left(I_{l}d_{l}I_{i}-I_{l}d_{i}I_{l}\right)P_e = i\hbar P_g\left(I_{i}I_{l}d_{l}-d_{i}I_{l}I_{l}\right)P_e - \hbar^2\epsilon_{ikl}I_k P_g d_l P_e\,.
\end{equation}
In vector form this reads:
\begin{align}
P_{g}\left(\mathbf{I}\times\mathbf{d}\right)P_e\left(\mathbf{I}\cdot\mathbf{J}\right)=i\hbar P_{g}\left[\mathbf{I}\left(\mathbf{I}\cdot\mathbf{d}\right)-\mathbf{I}^{2}\mathbf{d}\right] P_e
-\hbar^{2}P_{g}\left(\mathbf{I}\times\mathbf{d}\right)P_e\,.
\end{align}
Therefore, the relation~\eqref{eq:K-equation-App-1} transforms to
\begin{align}
\mathbf{K} =\frac{1}{\Delta}P_{g}\left|d_{ge}^{2}\right|\mathcal{\boldsymbol{\mathcal{E}}}
+
i\frac{A_{{\rm HF}}}{\Delta}
P_{g}
\left[
\left(1-\gamma\right)
\left(\mathbf{I}\times\mathbf{d}\right)
+i\frac{\gamma}{\hbar}
\left(\mathbf{I}\left(\mathbf{I}\cdot\mathbf{d}\right)-\mathbf{I}^{2}\mathbf{d}\right)
\right]P_e
GP_{e}\left(\mathbf{d}\cdot\boldsymbol{\mathcal{E}}\right)P_{g}\,.
\label{eq:K-equation-App-2}
\end{align}
The final step is to form the operator $\mathbf{K}$ from $\mathbf{d}$ and $G$ in Eq.~\eqref{eq:K-equation-App-2}. Using Eqs.~\eqref{eq:F_i_app}
and~\eqref{eq:D_sq_def-app},
one can write
\begin{equation}
P_{g}d_{i}GP_{e}\left(\mathbf{d}\cdot\boldsymbol{\mathcal{E}}\right)P_{g}=P_{g}d_{i}P_e GP_{e}d_{u}P_{g}\mathcal{E}_{u}=D_{iu}\mathcal{E}_{u}=K_{i}\,.
\end{equation}
Equivalently in the vector form one has
\begin{equation}
P_{g}\mathbf{d}P_e GP_{e}\left(\mathbf{d}\cdot\boldsymbol{\mathcal{E}}\right)P_{g}=\mathbf{K}\,.\label{eq:Relation-App-K-1}
\end{equation}
Analogously, the following relations hold:
\begin{equation}
P_{g}\left(\mathbf{I}\times\mathbf{d}\right)P_e GP_{e}\left(\mathbf{d}\cdot\boldsymbol{\mathcal{E}}\right)P_{g}=\textbf{\ensuremath{\mathbf{I}}}\times\mathbf{K},\label{eq:Relation-App-F-1}
\end{equation}
\begin{equation}
P_{g}(\mathbf{I}\cdot\mathbf{d})P_e GP_{e}\left(\mathbf{d}\cdot\boldsymbol{\mathcal{E}}\right)P_{g}=\mathbf{I}\cdot\mathbf{K}\,.
\end{equation}
Hence the relation~\eqref{eq:K-equation-App-2} provides the required equation for $\mathbf{K}$:
\begin{align}
\eta\mathbf{K} & =\frac{1}{\Delta}P_{g}\left|d_{ge}^{2}\right|\boldsymbol{\mathcal{E}}+i\frac{A_{{\rm HF}}}{\Delta}\left(1-\gamma\right)\left(\textbf{\ensuremath{\mathbf{I}}}\times\mathbf{K}\right)-\gamma\frac{A_{{\rm HF}}}{\hbar\Delta}\mathbf{I}\left(\textbf{\ensuremath{\mathbf{I}}}\cdot\mathbf{K}\right)\,,\label{eq:K-vector-equation-2-App}
\end{align}
where
\begin{equation}
\eta=1-\gamma A_{{\rm HF}}\hbar i_{I}\left(i_{I}+1\right)/\Delta\,.
\label{eta-appendix}
\end{equation}

\subsection{Solution of equation for $\mathbf{K}$}
\label{Subsec:Solution of equation for K}

As explained in the main text, we are looking for the solution of Eq.~\eqref{eq:K-vector-equation-2-App} in the form 
\begin{align}
\mathbf{K} & =\left[a_{0}\boldsymbol{\mathcal{E}}-i\frac{a_{1}}{\hbar}\textbf{\ensuremath{\mathbf{I}}}\times\boldsymbol{\mathcal{E}}+\frac{a_{2}}{\hbar^{2}}\textbf{\ensuremath{\mathbf{I}}}\left(\textbf{\ensuremath{\mathbf{I}}}\cdot\boldsymbol{\mathcal{E}}\right)\right]P_{g}\,,
\label{eq:K-vector-ansatz-1-App}
\end{align}

Substituting this ansatz into Eq.~\eqref{eq:K-vector-equation-2-App},
one finds
\begin{align}
\eta\mathbf{K} & =\frac{1}{\Delta}P_{g}\left|d_{ge}^{2}\right|\boldsymbol{\mathcal{E}}+i\frac{A_{{\rm HF}}}{\Delta}\left(1-\gamma\right)P_g\left[a_{0}\mathbf{I}\times\boldsymbol{\mathcal{E}}-i\frac{a_{1}}{\hbar}\mathbf{I}\times\left(\mathbf{I}\times\boldsymbol{\boldsymbol{\mathcal{E}}}\right)+\frac{a_{2}}{\hbar^{2}}\mathbf{I}\times\mathbf{I}\left(\mathbf{I}\cdot\boldsymbol{\mathcal{E}}\right)\right]+\mathbf{K}_{\gamma}\,,\label{eq:K-vector-eq-for-solution-2}
\end{align}
where 
\begin{equation}
\mathbf{K}_{\gamma}=-\gamma\frac{A_{{\rm HF}}}{\hbar\Delta}\mathbf{I}\left(\mathbf{I}\cdot\mathbf{K}\right)=-\gamma\frac{A_{{\rm HF}}}{\hbar\Delta}P_g\mathbf{I}\left[a_{0}\left(\mathbf{I}\cdot\boldsymbol{\mathcal{E}}\right)-i\frac{a_{1}}{\hbar}\mathbf{I}\cdot\left(\mathbf{I}\times\boldsymbol{\mathcal{E}}\right)+\frac{a_{2}}{\hbar^{2}}\left(\boldsymbol{\mathbf{I}}\cdot\boldsymbol{\mathbf{I}}\right)\left(\mathbf{I}\cdot\boldsymbol{\mathcal{E}}\right)\right]\,.
\end{equation}
The above relation can be rewritten as
\begin{equation}
\mathbf{K}_{\gamma}=-\gamma\frac{A_{{\rm HF}}}{\hbar\Delta}P_g\left[a_{0}+a_{1}+a_{2}\frac{\mathbf{I}^{2}}{\hbar^{2}}\right]\mathbf{I}\left(\mathbf{I}\cdot\boldsymbol{\mathcal{E}}\right)\,,
\label{K_gamma}
\end{equation}
where we used cyclic commutation relations for the components of the nuclear spin operator ${\bf I}$, that gives 
$\textbf{\ensuremath{\mathbf{I}}}\times\textbf{\ensuremath{\mathbf{I}}}=i\hbar\textbf{\ensuremath{\mathbf{I}}}$
(a general property of quantum angular momenta operators),
and also the relation 
$\mathbf{I}\cdot\left(\mathbf{I}\times\boldsymbol{\mathcal{E}}\right)=\boldsymbol{\mathcal{E}}\cdot\left(\boldsymbol{\mathbf{I}}\times\mathbf{I}\right)=i\hbar\boldsymbol{\mathbf{I}\cdot\boldsymbol{\mathcal{E}}}$.

Rearranging the vector operators entering the r.h.s. of Eq.~\eqref{eq:K-vector-eq-for-solution-2},
one has
\begin{align}
\eta\mathbf{K} & =\frac{1}{\Delta}P_{g}\left|d_{ge}^{2}\right|\boldsymbol{\mathcal{E}}+i\frac{A_{{\rm HF}}}{\Delta}\left(1-\gamma\right)P_g\left[\left(a_{0}+a_{1}\right)\boldsymbol{\mathbf{I}}\times\boldsymbol{\mathcal{E}}+\frac{i}{\hbar}\left(-a_{1}+a_{2}\right)\boldsymbol{\mathbf{I}}\left(\boldsymbol{\mathbf{I}}\cdot\boldsymbol{\mathcal{E}}\right)+i\frac{a_{1}}{\hbar}\boldsymbol{\mathcal{E}}\mathbf{I}^{2}\right]+\mathbf{K}_{\gamma}\,,\label{eq:K-vector-eq-for-solution-3}
\end{align}
where in the term proportional to $a_2$ relation $\textbf{\ensuremath{\mathbf{I}}}\times\textbf{\ensuremath{\mathbf{I}}}=i\hbar\textbf{\ensuremath{\mathbf{I}}}$ was applied, while in the term proportional to $a_1$ the calculation is as follows:
\begin{equation}
\big(\mathbf{I}\times(\mathbf{I}\times\boldsymbol{\mathcal{E}})\big)_k =
\epsilon_{kij}\epsilon_{mnj} I_i I_m \mathcal{E}_n =
I_i\mathcal{E}_i I_k -I_i I_i \mathcal{E}_k =
I_k I_i\mathcal{E}_i - I_i I_i \mathcal{E}_k + i\hbar\epsilon_{ikl}\mathcal{E}_i I_l =
I_k(\mathbf{I}\cdot\boldsymbol{\mathcal{E}})-\mathcal{E}_k\mathbf{I}^2+ i\hbar(\mathbf{I}\times\boldsymbol{\mathcal{E}})_k \,.
\end{equation}

Putting to Eq.~\eqref{eq:K-vector-eq-for-solution-3} the expression \eqref{K_gamma} for $\mathbf{K}_\gamma$ and grouping terms proportional to $\boldsymbol{\mathcal{E}}$, $\mathbf{I}\times\boldsymbol{\mathcal{E}}$ and $\mathbf{I}(\mathbf{I}\cdot\boldsymbol{\mathcal{E}})$, the following equation is obtained
\begin{align}
\eta\Delta\cdot\mathbf{K} & =\left[\left|d_{ge}^{2}\right|-A_{{\rm HF}}\left(1-\gamma\right)\mathbf{I}^{2}\frac{a_{1}}{\hbar}\right]\boldsymbol{\mathcal{E}}\,P_g
+\frac{i}{\hbar}\Big[\hbar A_{{\rm HF}}\left(1-\gamma\right)\left(a_{0}+a_{1}\right)\Big]\boldsymbol{\mathbf{I}}\times\boldsymbol{\mathcal{E}}\,P_g+\nonumber\\[10pt]
&\phantom{=}\ +\frac{1}{\hbar^2}\left[A_{\mathrm{HF}}\hbar\,\bigg(\left(1-\gamma\right)\left(a_{1}-a_{2}\right)-\gamma\,\Big(a_{0}+a_{1}+\tfrac{\mathbf{I}^{2}}{\hbar^{2}}a_{2}\Big)\bigg)\right]\boldsymbol{\mathbf{I}}\left(\boldsymbol{\mathbf{I}}\cdot\boldsymbol{\mathcal{E}}\right)P_g\,.\label{eq:K-vector-eq-for-solution-4}
\end{align}
Substituting Eq.~\eqref{eq:K-vector-ansatz-1-App} into \eqref{eq:K-vector-eq-for-solution-4} and comparing the factors at terms proportional to $\boldsymbol{\mathcal{E}}$, $\mathbf{I}\times\boldsymbol{\mathcal{E}}$ and $\mathbf{I}(\mathbf{I}\cdot\boldsymbol{\mathcal{E}})$,
one arrives at the following closed set of equations for the coefficients $a_{0}$, $a_{1}$ and $a_{2}$:
\begin{align}
\eta\Delta a_{0}&=\left|d_{ge}^{2}\right|-A_{{\rm HF}}\left(1-\gamma\right)\mathbf{I}^{2}\frac{a_{1}}{\hbar}\,,\label{eq:c_0-eq-gamma}\\
\eta\Delta a_{1}&=-\hbar A_{{\rm HF}}\left(1-\gamma\right)\left(a_{0}+a_{1}\right)\,,\\
\eta\Delta a_{2}&=\hbar A_{\mathrm{HF}}\left[\left(1-\gamma\right)\left(a_{1}-a_{2}\right)-\gamma\left(a_{0}+a_{1}+\frac{\mathbf{I}^{2}}{\hbar^{2}}a_{2}\right)\right]\,.
\end{align}
The solution of this system of equations reads
\begin{align}
a_{0}=&\frac{\left|d_{ge}^{2}\right|\left(\Delta+\hbar A_{\mathrm{HF}}\left(1-\gamma\left(\mathbf{I}^{2}/\hbar^{2}+1\right)\right)\right)}{\left(\Delta-E_{i_{I}+1}\right)\left(\Delta-E_{i_{I}-1}\right)}\,,\label{a_0-App}\\[7pt]
a_{1}=&-\frac{\hbar A_{\mathrm{HF}}\left|d_{ge}^{2}\right|\left(1-\gamma\right)}{\left(\Delta-E_{i_{I}+1}\right)\left(\Delta-E_{i_{I}-1}\right)}\,,\label{a_1-App}\\[7pt]
a_{2}=&-\frac{\hbar A_{\mathrm{HF}}\left|d_{ge}^{2}\right|\left[\hbar A_{\mathrm{HF}}+\gamma\left(\Delta-2\hbar A_{\mathrm{HF}}\right)-\gamma^{2}\hbar A_{\mathrm{HF}}\left(\mathbf{I}^{2}/\hbar^{2}-1\right)\right]}{\left(\Delta-E_{i_{I}+1}\right)\left(\Delta-E_{i_{I}}\right)\left(\Delta-E_{i_{I}-1}\right)}\,,\label{a_2-App}
\end{align}
where $E_{i_{I}}$ and $E_{i_{I}\pm 1}$ featured in denominators are eigenenergies given by Eqs.~\eqref{eq:E_i_plus_with-quadr-term}-\eqref{eq:E_i_minus_with-quadr-term} The numerators can also be represented in terms of eigenenergies, as in Eqs.~(\ref{eq:c_0-result-with-quadr-term})-(\ref{eq:c_2-result-with-quadr-term}) of the main text.

\section{A Clebsch-Gordan approach for calculating effective Hamiltonian}
\label{app:Clebsch-Gordan}

Throughout this paper, we used the Dyson equation
method for calculating the Green operator and thus the effective Hamiltonian.
The method has the advantage of being analytically solvable and providing a
physical intuition. Here, we summarize another method, a commonly used
Clebsch-Gordan approach which works well for numerical analysis but
is not convenient for analytics. The calculations based on the Clebsch-Gordan approach will
then be compared with the ones based on the Dyson equation approach.

We are interested in the effective Hamiltonian $\hat{H}_{\mathrm{eff}}$ describing
the light-induced coupling between the atomic nuclear spin states
in the ground electronic state manifold. It is given by Eqs.~\eqref{eq:G-definition}-\eqref{eq:D_ij} in
the main text, i.e. 
\begin{equation}
\hat{H}_{\mathrm{eff}}=\frac{1}{4}\mathcal{E}_{s}^{*}D_{s,q}\mathcal{E}_{q}\,,\label{eq:H_eff-def}
\end{equation}
where:
\begin{equation}
D_{s,q}=P_{g}d_{s}P_{e}GP_{e}d_{q}P_{g}\,,\quad\mathrm{with}\quad G=\left(\Delta-H_{{\rm HF}}\right)^{-1}\,.\label{eq:D_ij-Append_C}
\end{equation}
Here $P_{g}$ and $P_{e}$ are unit projector operators onto the
ground state ($^{1}S_{0}$) and excited state ($^{3}P_{1}$) manifolds, respectively. In the latter $^{3}P_{1}$ manifold the eigenstates of
the hyperfine Hamiltonian $H_{{\rm HF}}$ are $|j_{e}\,i_{I}\,f\,m_{f}\rangle$,
where $j_{e}=1$ is electron total angluar momentum and $i_{I}-1\le f\le i_{I}+1$
is the total angular momentum of the atom. The corresponding hyperfine energies
$E_{f}$ are given by Eqs.~\eqref{eq:E_i_plus_with-quadr-term} and
\eqref{eq:E_i_minus_with-quadr-term} in the main text. The Green
operator projected onto this excited state manifold is diagonal in
the basis of eigenstates $|j_{e}\,i_{I}\,f\,m_{f}\rangle$ and reads explicitly: 

\begin{equation}
P_{e}GP_{e}=\sum_{f=i_{I}-j_{e}}^{i_{I}+j_{e}}\sum_{m_{f}=-f}^{f}\frac{|j_{e}\,i_{I}\,f\,m_{f}\rangle\langle j_{e}\,i_{I}\,f\,m_{f}|}{\Delta-E_{f}}\,.
\end{equation}
Expressing hyperfine eigenstates $|j_{e}\,i_{I}\,f\,m_{f}\rangle$
in terms of the bare electron and spin states $|j_{e}\,i_{I}\,m_{j_{e}}\,m_{i}\rangle \equiv |j_{e}\,m_{j_{e}}\rangle|i_{I}\,m_{i}\rangle$
via the Clebsch-Gordan coefficients $C_{j_{e}i_{I}m_{j_{e}}m_{i}}^{fm_{f}}$,
one has 

\begin{equation}
P_{e}GP_{e}=\sum_{m_{j},m_{i},m_{j}^{\prime},m_{i}^{\prime},f,m_{f}}|j_{e}\,m_{j_{e}}\rangle|i_{I}\,m_{i}\rangle\frac{C_{j_{e}i_{I}m_{j_{e}}m_{i}}^{fm_{f}}C_{j_{e}i_{I}m_{j_{e}}^{\prime}m_{i}^{\prime}}^{fm_{f}*}}{\Delta-E_{f}}\langle j_{e}\,m_{j_{e}}^{\prime}|\langle i_{I}\,m_{i}^{\prime}|\,.
\end{equation}

Finally, since the dipole operator acts only on the electronic degrees
of freedom, Eq.~(\ref{eq:D_ij-Append_C}) for the tensor $D_{s,q}$
takes the form:
\begin{equation}
D_{s,q}=P_{g}\sum_{m_{i},m_{i}^{\prime}}|i_{I}\,m_{i}\rangle D_{s,q}^{m_{i},m_{i}^{\prime}}\langle i_{I}\,m_{i}^{\prime}|\,,\label{eq:D_ij-Clebsch-final}
\end{equation}
with
\begin{equation}
D_{s,q}^{m_{i},m_{i}^{\prime}}=\sum_{m_{j},m_{j}^{\prime},f,m_{f}}\langle j_{g}\,m_{j_{g}}|d_{s}|j_{e}\,m_{j}\rangle\frac{C_{j_{e}i_{I}m_{j_{e}}m_{i}}^{fm_{f}}C_{j_{e}i_{I}m_{j_{e}}^{\prime}m_{i}^{\prime}}^{fm_{f}*}}{\Delta-E_{f}}\langle j_{e}\,m_{j_{e}}^{\prime}|d_{q}|j_{g}\,m_{j_{g}}^{\prime}\rangle\,,\label{eq:D_ij-Clebsch-final^m_i,m'_i}
\end{equation}
where $j_{g}=0$ and $m_{j_{g}}=0$ are, respectively, the electron angular momentum
and its projection quantum numbers for the ground state
manifold $^{1}S_{0}$, and
the unit projection operator onto this manifold is $P_{g}=|j_{g}\,m_{j_{g}}\rangle\langle j_{g}\,m_{j_{g}}|$.

We now define the dipole moment matrix elements $\langle j_{g}\,m_{j_{g}}|d_{s}|j_{e}\,m_{j}\rangle\equiv\langle^{1}S_{0}|d_{s}|{}^{3}P_{1},m_{j}\rangle$
featured in Eq.~(\ref{eq:D_ij-Clebsch-final^m_i,m'_i}). As discussed
in ref. \cite{LeBellac2003} (chapt.~3 p.~50, chapt.~10 p.~350-351), if
we identify the two-photon polarisation states $\ket{\sigma^{\pm}}$
to photon angular momentum states $\ket{j=1,m=\pm1}$, and defining
the photon angular momentum operator $\vec{J}$, the states must transform
as $\exp{-i\pi J_{y}}\ket{\sigma^{+}}=+\ket{\sigma^{-}}$ to satisfy
the standard raising and lowering operator conventions. The commonly
used relations between circular and linear polarization states $\ket{\sigma^{\pm}}=(\ket{x}\pm i\ket{y})/\sqrt{2}$
do not satisfy that, and will lead to erroneous tensorial shift calculations.
Instead one must use $\ket{\sigma^{\pm}}=(\mp\ket{x}-i\ket{y})/\sqrt{2}$.
Consequently, the matrix elements of the atomic dipole operators $d_{x,y,z}=\vec{d}\cdot\vec{e}_{x,y,z}$
are

\begin{equation}
\langle^{1}S_{0}|d_{x}|{}^{3}P_{1},m_{j}\rangle=\frac{d_{ge}}{\sqrt{2}}\left(\delta_{m_{j},-1}-\delta_{m_{j},1}\right)\,,\label{eq:d_x-element}
\end{equation}
\begin{equation}
\langle^{1}S_{0}|d_{y}|{}^{3}P_{1},m_{j}\rangle=-\frac{d_{ge}}{i\sqrt{2}}\left(\delta_{m_{j},1}+\delta_{m_{j},-1}\right)\,,\label{eq:d_y-element}
\end{equation}
\begin{equation}
\langle^{1}S_{0}|d_{z}|{}^{3}P_{1},m_{j}\rangle=d_{ge}\delta_{m_{j},0}\,.\label{eq:d_z-element}
\end{equation}
Using Eqs.~(\ref{eq:H_eff-def}), (\ref{eq:D_ij-Clebsch-final})-(\ref{eq:D_ij-Clebsch-final^m_i,m'_i})
and (\ref{eq:d_x-element})-(\ref{eq:d_z-element}), the matrix elements
of the effective Hamiltonian were numerically calculated and compared
with analytical results in Fig.~\ref{fig:fig2} showing a perfect agreement
between numerical and analytical results.

\end{widetext}

\bibliography{biblio}

\end{document}